\documentclass[a4paper,11pt]{article}
\pdfoutput=1 

\usepackage{amssymb}
\usepackage{amsmath}
\usepackage{amsfonts}
\usepackage{cite}
\usepackage{epsfig}
\usepackage{graphicx}
\usepackage{lineno}
\usepackage{subcaption}
\usepackage{bbm}
\usepackage{dsfont}

\usepackage{simpler-wick}
\usepackage{tikz-feynman}
\tikzfeynmanset{compat=1.1.0}
\tikzfeynmanset{warn luatex=false}
\usepackage{tikz,environ}
\usetikzlibrary{decorations}
\usetikzlibrary{arrows}
\usetikzlibrary{decorations.markings}
\tikzset{
	hard/.style={postaction={decorate},
		line width=0.5mm
	},
		soft/.style={postaction={decorate},
		line width=0.5mm, dashed
	},
	momentum/.style={postaction={decorate},
	line width=0.5mm,
	color=gray,
	decoration={
    markings,
    mark=at position 0.8 with {\arrow{stealth}}}
    },
	hardarrow/.style={postaction={decorate},
	line width=0.5mm,
	decoration={
    markings,
    mark=at position 0.6 with {\arrow{stealth}}}},
}

\newcommand{\cA}{\mathcal{A}}

\newcommand{\cO}{\mathcal{O}}

\newcommand{\cL}{\mathcal{L}}

\newcommand{\tr}{\mathrm{Tr}}
\newcommand{\nn}{\nonumber}
\newcommand{\emCoupling}{g}
\newcommand{\photonField}{A}

\newcommand{\e}{\varepsilon}

\newcommand{\jA}{\mathcal{J}}

\newcommand{\jAbar}{{{\mathcal{J}}_{H}}}

\newcommand{\angMom}{L}

\newcommand{\constK}{\mathcal K}

\DeclareMathOperator*{\sumint}{%
\mathchoice%
  {\ooalign{$\displaystyle\sum$\cr\hidewidth$\displaystyle\int$\hidewidth\cr}}
  {\ooalign{\raisebox{.14\height}{\scalebox{.7}{$\textstyle\sum$}}\cr\hidewidth$\textstyle\int$\hidewidth\cr}}
  {\ooalign{\raisebox{.2\height}{\scalebox{.6}{$\scriptstyle\sum$}}\cr$\scriptstyle\int$\cr}}
  {\ooalign{\raisebox{.2\height}{\scalebox{.6}{$\scriptstyle\sum$}}\cr$\scriptstyle\int$\cr}}
}

\usepackage{jheppub} 
\usepackage[T1]{fontenc} 

\usepackage{cleveref}
\crefname{figure}{Fig.}{Figs.}
\Crefname{figure}{Fig.}{Figs.}
\crefname{equation}{Eq.}{Eqs.}
\Crefname{equation}{Eq.}{Eqs.}
\crefname{section}{Section}{Sections}
\Crefname{Section}{Section}{Sections}

\title{\boldmath Soft Theorems from Higher Symmetries}

\author[a,b]{Jonah Berean-Dutcher}
\author[c]{Maria Derda,}
\author[b]{Julio Parra-Martinez}

\affiliation[a]{University of British Columbia, Vancouver, V6T 1Z1, Canada}
\affiliation[b]{Institut des Hautes Études Scientifiques, 91440 Bures-sur-Yvette, France}
\affiliation[c]{Walter Burke Institute for Theoretical Physics, Caltech, Pasadena, California 91125, USA}

\emailAdd{jbd@phas.ubc.ca}
\emailAdd{mderda@caltech.edu}
\emailAdd{julio@ihes.fr}

\abstract{We describe the connection between spontaneously-broken higher symmetries and soft theorems for scattering amplitudes of their associated Nambu-Goldstone bosons. Our main result is a new sub-leading double soft pion theorem in theories with a spontaneously-broken continuous 2-group global symmetry, which intertwines amplitudes with different numbers of pions and photons. We also provide a novel derivation of the leading soft photon theorem from the Ward identity of an emergent 1-form global symmetry in effective field theories where antiparticles are integrated out. Our derivations of these soft theorems use the algebra of spacetime currents and do not rely on asymptotic symmetries or diagrammatic arguments.}

\begin{document} 
\maketitle
\flushbottom
\newpage
\section{Introduction}

Symmetry is one of the cornerstones of our understanding of nature at all scales, from the world of fundamental particles to the physics of phase transitions and the early universe. Perhaps the most striking consequence of global symmetry in quantum field theory (QFT) is the existence of massless scalar particles, predicted by Nambu \cite{Nambu:1960tm} and Goldstone \cite{Goldstone:1961eq, Goldstone:1962es},  when such symmetry is continuous and spontaneously broken. Beyond their existence, the dynamics of such Nambu-Goldstone bosons (NGB) is weakly coupled and heavily constrained by the broken symmetry. This allows a detailed effective field theory (EFT) description of their long-distance interactions, which lets us understand a variety of systems: from ferromagnetic materials to the dynamics of pions in low-energy quantum chromodynamics (QCD) \cite{Weinberg:1978kz, Manohar:2018aog, Burgess:2020tbq}.

A direct implication of spontaneously broken symmetry in the dynamics of NGB are \emph{soft theorems}, which govern the universal behavior of the scattering amplitudes of NGB in the limit in which some of their momenta are taken to zero. An early example is the so-called Adler zero \cite{Adler:1964um} in the low-energy dynamics of pions, and its generalizations \cite{Weinberg:1966gjf,Dashen:1969eg} including to multi-soft pion limits, \cite{Weinberg:1966kf, Arkani-Hamed:2008owk,Kampf:2013vha,Distler:2018rwu,Low:2015ogb}, subleading \cite{Cachazo:2015ksa,Low:2015ogb,Du:2015esa}, higher-derivative corrections \cite{Rodina:2021isd}, and other theories of NGB \cite{Alonso:2016oah,Kampf:2019mcd,Green:2022slj,DiVecchia:2015jaq,Cheung:2023qwn}.
    Soft limits are not limited to scalar NGB, and are found to be universal for other massless particles, including scalar moduli \cite{Cheung:2021yog,Cheung:2022vnd,Derda:2024jvo, Cohen:2025dex}, photons \cite{Low:1958sn,Weinberg:1964ew,Weinberg:1965nx,Burnett:1967km}, gluons \cite{Bassetto:1983mvz,Berends:1988zn,Casali:2014xpa} and gravitons \cite{Weinberg:1964ew,Weinberg:1965nx,Cachazo:2014fwa,Bern:2014vva,DiVecchia:2015jaq}. However, our traditional description of such soft theorems often requires going beyond the language of spontaneous symmetry breaking or ordinary symmetries, and utilizes notions of asymptotic symmetries \cite{Strominger:2017zoo,Strominger:2013jfa,Kapec:2015vwa,Campiglia:2015qka,Himwich:2019dug,He:2019jjk,Kapec:2021eug,Kapec:2022axw,Kapec:2022hih,Adamo:2023zeh,Chen:2023tvj,Agrawal:2025bsy}, or other considerations \cite{Bern:2014vva,Larkoski:2014bxa}.

Recently, there has been a leap in our understanding of what constitutes a symmetry in QFT. Beyond the familiar textbook symmetries that act on local operators and whose charges are carried by particles, there exists a zoo of so-called \emph{generalized symmetries} or \emph{higher symmetries} (for references, see \cite{Cordova:2022ruw}). These include higher-form symmetries, which act on lower-codimension operators (lines, surfaces, etc.), and are carried by extended excitations (strings, membranes, etc.), higher-group symmetries \cite{Kapustin:2013uxa,Sharpe:2015mja,Tachikawa:2017gyf,Cordova:2018cvg,Benini:2018reh,Baez:2004in,Brennan:2020ehu,Arbalestrier:2025poq,Cordova:2022qtz}, which mix symmetries of various ranks, and even non-invertible symmetries \cite{Komargodski:2020mxz,Kaidi:2021xfk,Choi:2021kmx,Cordova:2022ieu,GarciaEtxebarria:2022jky,Choi:2022jqy}. A simple example of these is the (electric) 1-form symmetry of Maxwell theory, which acts on Wilson lines and has as current the electromagnetic field-strength, conserved by the vacuum Maxwell's equations.
From this modern perspective, non-scalar massless particles, such as the photon, can often be interpreted as the NBG (or pseudo-NGB) associated with the higher symmetry \cite{Gaiotto:2014kfa, Hofman:2018lfz,Kovner:1992pu}. These novel symmetries have found wide application in high-energy and condensed-matter physics \cite{Cordova:2022ruw,McGreevy:2022oyu}, but their consequences on the on-shell S-matrix describing scattering processes remain largely unexplored (with a few notable exceptions \cite{Copetti:2024rqj,Copetti:2024dcz,vanBeest:2023dbu,vanBeest:2023mbs}).

    In this paper we argue that these new notions of symmetry enable the unification of our understanding of soft theorems for the various kinds of NGB.
 In particular, we will explain that the familiar soft photon theorem can be derived as a consequence of 1-form symmetries which act on line operators and as a shift symmetry on the electromagnetic field. Such symmetry is emergent in the limit where particle production is suppressed, which can be made precise using the language of heavy-particle effective theories such as Heavy Quark EFT (HQET)  \cite{Manohar:2000dt, Damgaard:2019lfh}. Furthermore we will derive a new double soft pion theorem, in theories with a spontaneously broken continuous 2-group symmetry, wherein the current algebra of ordinary (or 0-form, in modern parlance) symmetry is intertwined with that of a 1-form symmetry. This double soft theorem includes contributions which change particle species---between NGB of the 0-form symmetry (pions) and of the 1-form symmetry (photons)---thereby reflecting the intertwined current algebra. As an example, we focus on the low-energy effective theory of QCD with massless quarks and with gauged $U(1)_V$ vector symmetry corresponding to Baryon number, which shows that such structure is far from exotic \cite{FileviezPerez:2010gw,FileviezPerez:2011pt,Duerr:2013dza,FileviezPerez:2014lnj,FileviezPerez:2024fzc,Butterworth:2024eyr,Butterworth:2025asm}.

The rest of this paper is organized as follows: in~\cref{sec:0-form} we review the familiar understanding of scalar NGB, which henceforth we call pions, and the derivation of their single- and double soft theorems from the current algebra of 0-form symmetry. In~\cref{sec:1-form} we review the modern perspective on the photon as a NGB and show that leading soft photon theorems follow from the Ward identities of emergent 1-form symmetries. Finally in \cref{sec:new-soft-theorems-from-2-group} we derive the new double soft pion theorem for theories with spontaneously broken continuous 2-group symmetry using its current algebra, and illustrate it with some examples at tree level. Finally, in an appendix we provide some technical details for the current algebra proof of the known sub-leading double soft pion theorem (which is presented here for the first time).

{\bfseries Conventions}:
This paper deals with scattering amplitudes, so we work in Lorentzian signature with a mostly-minus metric \(\left(+\,,\,-\,,\,-\,,\,-\right)\). This is in contrast to most of the literature on generalized symmetries which uses Euclidean signature. Our scattering amplitudes are defined with all external momenta taken to be outgoing.


\section{Review: Soft pions from ordinary symmetry}
\label{sec:0-form}

In this section we will review the spontaneous breaking of continuous 0-form symmetry and the associated soft theorems for the scattering amplitudes of pions. This is all textbook material, which we choose to collect  here with the purpose of highlighting the similarities with our discussion of 1-form symmetry in Section~\ref{sec:1-form}, and to provide the background for our new double soft theorem in Section~\ref{sec:new-soft-theorems-from-2-group}. The only novelty here is the current algebra derivation of the known sub-leading double soft pion theorem,\footnote{For the case of spontaneously broken space-time symmetries, see \cite{Guerrieri:2017ujb} for a current algebra derivation of the corresponding leading and sub-leading double soft theorems.} whose technical details are collected in Appendix~\ref{app:select-form-factors-0-form}.

Continuous 0-form symmetries have conserved currents 
\begin{equation}
    \partial^\mu j_{\mu}^a(x)=0\,,
\end{equation}
which we can integrate over codimension-1 cycles, $\Sigma_{D-1}$ to define an associated conserved charges:
\begin{equation}
    Q^a(\Sigma_{D-1}) = \int_{\Sigma_{D-1}} d^{d-1}x \, \hat{n}^\mu   j^a_{\mu} \,.\label{eq:0formQ}
\end{equation}
Here $\hat{n}$ is a unit normal vector to $\Sigma_{D-1}$. These conserved charges are in one-to-one correspondence with the Lie algebra generators of $G$, $t^a$, satisfying
\begin{align}
    [t^a,t^b] = if^{abc}t^c\,.
\end{align}
Local operators may be charged under 0-form symmetries. Given a 0-form symmetry with group $G$, an operator transforming under $G$ with representation $\mathcal{R}$ will satisfy the Ward identity,
\begin{align}
     \partial^\mu j^a_\mu(x)\mathcal{O}_\mathcal{R}(y) &= \delta^{(4)}(x-y) t_{\mathcal{R}}^a\mathcal{O}_\mathcal R(y)\,,
     \label{eq:nlsm-operator-eqns-general-operator}
\end{align}
where $t^a$ are the generators of $G$.  If such a local operator develops a vacuum expectation value
\begin{equation}
    \langle \mathcal{O}_\mathcal R(y) \rangle \sim v^a\,.\label{eq:O-vev}
\end{equation}
the symmetry is spontaneously broken down to a subgroup $H$ which is generated by the charges $\{T^a\} \subset \{t_a\}$ that leave the vacuum expectation value invariant $[v^a,T^a] = 0$. The rest of the generators, ${X^a}$, form a coset $G/H$. This implies the following decomposition of the commutation relations for the generators of $G$,
\begin{align}
    [T^a,T^b] &= i f_T^{abc}T^c\,,\\
    [X^a,X^b] &= i F^{abc}T^c\,,\\
    [T^a,X^b] &= i f_X^{abc}X^c\,.
\label{eq:nlsm-lie-algebra}
\end{align}
The $f_T^{abc}$, $f_X^{abc}$, and $F^{abc}$ are structure constants which satisfy the appropriate Jacobi identities, and coincide in the case of the chiral symmetry breaking in QCD. Here we have assumed that the coset $G/H$ is a symmetric space, i.e., that it has the additional symmetry $X^a\to - X^a$, sometimes called $G$-parity.

Goldstone's theorem \cite{Goldstone:1962es} implies that the spectrum contains a massless NGB -- a pion $\pi^a$ -- for every broken generator $X^a$. More concretely, there exists conserved currents $\jA_\mu^a(x)$ corresponding to the broken generators, which interpolate the pion,
\begin{align}
    \langle \pi^b(p) | \jA_\mu^a(x)\rangle = if_\pi p_\mu \delta^{ab}e^{ip\cdot x}. \label{eq:goldstone-interpolation}
\end{align}
Here $f_\pi$ is the dimensionful pion decay constant. The interpolation property implies that this current contains a contribution that is linear in the pion field,
\begin{align}
    \jA_\mu^a(x) = f_\pi \partial_\mu\pi^a(x) + \mathcal{O}(\pi^3)\,.
\end{align}
Thus, the symmetry is non-linearly realized and at leading order acts as a constant shift of the pion field
\begin{align}
    \pi^a(x) \rightarrow \pi^a(x) + c^a  + \dotsm\,,  \qquad \text{with} \qquad \partial_\mu c^a =0\,. \label{eq:pion-shift} 
\end{align}
In contrast, the $\pi^a$ transform as the broken generators, that is, in a linear representation of the unbroken subgroup $H$, which infinitesimally takes
\begin{align}
    \pi^a(x) \rightarrow \pi^a(x) + f^{abc}_X\pi^b(x)\lambda^c\,.
\end{align}
The associated symmetry current then has the form 
\begin{align}
    V_\mu^a(x) = -f_X^{abc} \pi^b(x) \partial_\mu \pi^c(x)+\mathcal{O}(\pi^4). 
\end{align}
The symmetry structure furnished by \cref{eq:nlsm-lie-algebra} implies the following set of operator equations for the currents,
\begin{align}
    \partial^\mu V^a_\mu(x)V_\nu^b(y) &= if_T^{abc} \delta^{(4)}(x-y)V^c_\nu(y) \label{eq:nlsm-current-algebra-VV}\\
    \partial^\mu \jA^a_\mu(x)\jA_\nu^b(y) &= iF^{abc} \delta^{(4)}(x-y)V^c_\nu(y) 
    \label{eq:nlsm-current-algebra-JJ}\\
        \partial^\mu \jA^a_\mu(x)V_\nu^b(y) &=  if_X^{abc}\delta^{(4)}(x-y) \jA^c_\nu(y)\,.
\label{eq:nlsm-current-algebra-JV}
\end{align}
From this point on we will focus on theories possessing a chiral 0-form global symmetry $G =G_L \times G_{R}$ which is spontaneously broken down to a vector subgroup $H=G_V$, so we will refer to $\jA^a_\mu(x)$ as the axial current, and to $V^a_\mu(x)$ as the vector current.  In the example of the chiral effective theory of pions describing (massless) QCD at low energies, $G=SU(N_f)_L\times SU(N_f)_R$ and is broken down to the vector subgroup, $SU(N_f)_V$. 

An effective theory for these pion NGB can be written in terms of the charged operator
\begin{align}
    U(x) = \mathrm{exp}\left(\frac{2i}{f_\pi}\pi(x)^a t^a\right), \label{eq:chiral-field}
\end{align}
parameterizing fluctuations of the vacuum expectation value in \Cref{eq:O-vev},
where we choose the normalization $\tr\left(t^a t^b\right) = \frac{1}{2}\delta^{ab}$. The effective Lagrangian for a general 0-form spontaneous symmetry breaking pattern $G_L \times G_{R}\rightarrow G_V$ is given by \cite{Coleman:1969sm,Callan:1969sn, Meetz:1969as,Volkov:1973vd,Honerkamp:1971xtx,Weinberg:1978kz,Boulware:1981ns}
\begin{align}
    \cL = \frac{f_\pi^2}{4}\tr\left(\partial_\mu U \partial^\mu U^{\dagger}\right) +\cdots\,.
    \label{eq:lagrangian-nlsm}
\end{align}
where the dots include higher derivative terms. This is the famous non-linear sigma model (NLSM). 

\subsection{Pion Adler zero}

It follows from \cref{eq:goldstone-interpolation} that momentum-space form factors describing the overlap of the axial current with an on-shell scattering state of $n$-many outgoing pions,  $|\alpha \rangle$, admit a decomposition
\begin{align}
    \langle \alpha | \jA^a_\mu(q)\rangle=&\frac{f_\pi q_\mu}{q^2}\langle \alpha+ \pi^a(q)|0 \rangle+\langle \alpha|\jAbar^a_\mu(q)\rangle\,. \label{eq:axial-current-decomposition}
\end{align}
The first term carries the single-pion pole, whose coefficient is the matrix element corresponding to the amplitude with a additional pion with momentum $q$
\begin{equation}
    \langle \alpha+ \pi^a(q)|0 \rangle = i \cA_{n+\pi(q)}\,.
\end{equation}
where we leave a momentum-conserving delta function implicit.
The second  term contains the `hard part of the current' $\jAbar^a_\mu$, which includes the infinite number of higher-order contributions that make up the axial current. The axial current is conserved in the correlation function on the LHS of the above equation since there are no other insertions of charged operators in the correlator. Conservation then implies the relation
\begin{align}
     \cA_{n+\pi(q)} = \frac{i}{f_\pi} q^\mu\langle \alpha|\jAbar^a_\mu(q)\rangle
     \,.
    \label{eq:axial-current-conservation}
\end{align}
Taking the limit $q\rightarrow0$ of this equation yields the soft theorem known as the Adler zero \cite{Adler:1964um},
\begin{align}
    \lim_{q\rightarrow 0} \cA_{n+\pi(q)} = 0 + \mathcal{O}(q)\,,
    \label{eq:adler-zero-nlsm}
\end{align}
where we have assumed that $\langle \alpha|\jAbar^a_\mu(q)\rangle$ is regular in this limit. This regularity follows from  spontaneous symmetry breaking, and is realized in the theory by the lack of any cubic coupling for the pions. 

In summary, we have seen that the Adler zero follows from the spontaneously broken symmetry as a consequence of Goldstone's theorem and the conservation of the axial current.

\subsection{Double soft pion theorem}
Now we will discuss the double soft pion theorem \cite{Weinberg:1966gjf,Weinberg:1966kf} which encodes the non-abelian current algebra described by the operator equations in \cref{eq:nlsm-current-algebra-VV,eq:nlsm-current-algebra-JJ,eq:nlsm-current-algebra-JV}. Several different derivations of the double soft theorem for pions exist in the literature, for example in \cite{Arkani-Hamed:2008owk,Distler:2018rwu}. The sub-leading contribution for flavor-dressed amplitudes was first derived in \cite{Cachazo:2015ksa,Low:2015ogb,Du:2015esa} at tree-level or using diagrammatic arguments. Here we follow Ref.~\cite{Kampf:2013vha} and extend the analysis to emphasize how the leading and sub-leading theorem are a direct consequence of current algebra.

The starting point is the momentum-space correlator with two axial current operator insertions, which satisfies the Ward identity \cref{eq:nlsm-current-algebra-JJ} in either of the currents. For the current with momentum $q_1$ we have
\begin{align}
    q_1^\mu  \langle\alpha| \jA^a_\mu(q_1)\jA^b_\nu(q_2) \rangle=
        F^{abc} \langle \alpha|V^c_\nu(q_1+q_2) \rangle\,,
    \label{eq:nlsm-JJ-soft}
\end{align}
and there is an analogous Ward identity for current with momentum $q_2$.
Hence we can write the RHS of \cref{eq:nlsm-JJ-soft} in the manifestly symmetric form
\begin{align}
    q_1^\mu q_2^\nu\langle\alpha| \jA^a_\mu(q_1)\jA^b_\nu(q_2) \rangle=-\frac{1}{2}F^{abc}(q_1-q_2)^\mu\langle \alpha|V^c_\mu(q_1+q_2) \rangle\,. \label{eq:double-soft-ward}
\end{align}
On the other hand, decomposing axial currents according to \cref{eq:axial-current-decomposition} yields 
\begin{align}
    q_1^\mu q_2^\nu\langle \alpha| \jA^a_\mu(q_1)\jA^b_\nu(q_2)\rangle=&f^2_\pi \langle \alpha+\pi^a(q_1)\pi^b(q_2)|0 \rangle+ f_\pi q_1^\mu \langle \alpha+ \pi^b(q_2)|\jAbar^a_\mu(q_1) \rangle\nonumber\\
    &+ f_\pi q_2^\nu\langle \alpha+\pi^a(q_1)| \jAbar^b_\nu(q_2)\rangle +q_1^\mu q_2^\nu\langle  \alpha|\jAbar^a_\mu(q_1)\jAbar^b_\nu(q_2)\rangle\,,
\end{align}
which with the help of \cref{eq:axial-current-conservation} we can rewrite as
\begin{align}
    q_1^\mu q_2^\nu\langle \alpha|\jA^a_\mu(q_1)\jA^b_\mu(q_2)\rangle= -f^2_\pi \langle\alpha+ \pi^a(q_1)\pi^b(q_2)|0 \rangle+q_1^\mu q_2^\nu\langle \alpha| \jAbar^a_\mu(q_1) \jAbar^b_\nu(q_2)\rangle. \label{eq:double-soft-decomposition}
\end{align}
The first term on the RHS of \cref{eq:double-soft-decomposition} contains a scattering amplitude of $(n+2)$-many pions. Combining \cref{eq:double-soft-ward} and \cref{eq:double-soft-decomposition} we obtain
\begin{align}
    &f^2_\pi \langle\alpha+ \pi^a(q_1)\pi^b(q_2)|0 \rangle = \frac{1}{2}F^{abc}(q_1-q_2)^\mu\langle \alpha| V^c_\mu(q_1+q_2)\rangle + q_1^\mu q_2^\nu\langle \alpha| \jAbar^a_\mu(q_1)\jAbar^b_\nu(q_2)\rangle\, . \label{eq:double-soft-correlators}
\end{align}
To arrive at the double soft theorem we take the simultaneous limit $q_1, q_2\rightarrow 0$ of this expression. In Appendix \ref{app:select-form-factors-0-form} we derive the applicable soft limits of the two form factors on the RHS which are shown to be related to $n$-point scattering amplitudes of the remaining $n$ hard pions $\pi^{a_i}(p_i)$ in the state $\langle \alpha |$.
More concretely we find that at sub-leading order the soft vector current form factor is given by 
\begin{align}
    \lim_{q\to0}\langle \alpha| V^c_\mu(q)\rangle  =-i\sum_{i=1}^n f^{a a_i d}\left(\frac{(2p_{i}+q)_\mu}{(p_i+ q)^2} -\frac{iq^\nu \angMom_{i\mu\nu}}{(p_i\cdot q)} \right)\cA_{n}^{a_1\ldots d\ldots a_n}\,,
\end{align}
where the angular momentum operator $\angMom_i^{\mu\nu}$ is defined as
\begin{align}
    \angMom_i^{\mu\nu} = i \left(p_{i}^{\mu}\frac{\partial}{\partial p_{i}^{\nu}}-p_{i}^{\nu}\frac{\partial}{\partial p_{i}^{\mu}}\right). \label{eq:angular-momentum-operator}
\end{align}
Meanwhile, the hard axial currents form factor at leading order in soft momenta is 
\begin{align}
    \lim_{q_1q_2\rightarrow0}\langle\alpha|  \jAbar^a_\mu(q_1)\jAbar^b_\nu(q_2)\rangle
    = -i\sum_{i=1}^n(F^{aa_ie}f_X^{ebd}+F^{ba_ie}f_X^{ead}) \frac{\eta_{\mu\nu}}{2p_i\cdot (q_1+q_2)}\cA^{a_1\cdots d\cdots a_n}_n\,.
\end{align}
Combining these results we obtain for an $(n+2)$-particle amplitude with soft particles $\pi^a( q_1)$ and $\pi^b(q_2)$ and $n$ hard particles $\pi^{a_i}(p_i)$,
\begin{align}
   \lim_{q_1,q_2\to0}\mathcal{A}_{n+\pi^a( q_1)\pi^b( q_2)} = \left(S^{(0)} + S^{(1)}\right) \mathcal{A}_{n}\, ,
   \label{eq:double_soft_nlsm}
\end{align}
where we scale both soft momenta to zero simultaneously. The leading soft factor is,
\begin{align}
S^{(0)}\mathcal{A}_{n}= -\frac{1}{f_\pi^2}\sum^{n}_{i=1}F^{abc}f_X^{ca_i d}\frac{p_i\cdot(q_1-q_2)}{2p_i\cdot(q_1+q_2)}\cA^{a_1\ldots d\ldots a_n}_{n}, \label{eq:double-soft-leading-order}
\end{align}
and the sub-leading  soft factor is\footnote{The sub-leading order soft factor is modified in the presence of four-derivative operators in the chiral Lagrangian; see \cite{Rodina:2021isd} for a full expression. It also receives loop corrections that will be explored elsewhere.}
\begin{align}
S^{(1)}\mathcal{A}_{n}=
&-\frac{1}{f_\pi^2}\sum^{n}_{i=1}(F^{aa_i c}f_X^{cb d}+F^{ba_i c}f_X^{ca d})\frac{(q_1 \cdot q_2)}{2p_i\cdot(q_1+q_2)}\cA^{a_1\ldots d\ldots a_n}_{n}\nonumber \\
&+\frac{1}{f_\pi^2}\sum^{n}_{i=1}F^{ab c}f_X^{ca_i d}\frac{(q_1 \cdot q_2)(p_i\cdot(q_1-q_2))}{2(p_i\cdot(q_1+q_2))^2}\cA^{a_1\ldots d\ldots a_n}_{n}\nonumber \\
&+\frac{1}{f_\pi^2}\sum^{n}_{i=1}F^{ab c}f_X^{ca_i d}\frac{iq_1^\mu q_2^\nu \angMom_{i\mu\nu}}{p_i\cdot(q_1+q_2)}\cA^{a_1\ldots d\ldots a_n}_{n}\,,
\label{eq:NLSM-double-soft-subleading-order}
\end{align}
with the angular momentum operator $\angMom_i^{\mu\nu}$ given by \cref{eq:angular-momentum-operator}.


\section{Soft photons from higher-form symmetry}
\label{sec:1-form}

In this section we will discuss photons as NGB for 1-form global symmetries. We will show how the development of the previous section for 0-form NGB is suitably generalized and how the familiar soft photon theorems derive from corresponding Ward identities. 

\subsection{Photons as Nambu-Goldstone bosons}

Let us begin by quickly reviewing some basic facts about continuous 1-form symmetries, and the modern perspective of the photon as a NGB. We aim to be pedagogical and to stress the similarities with our discussion of 0-form symmetry breaking in the previous section. This discussion can (and probably should) be skipped by the experts.

Continuous 1-form symmetries have conserved currents which are antisymmetric rank-2 tensors
\begin{equation}
    J_{\mu\nu}(x) = J_{[\mu\nu]}(x)\,, \qquad \text{with} \qquad \partial^\mu J_{\mu\nu}(x)=0\,.
\end{equation}
The corresponding conserved charges are integrals over codimension-2 cycles, $\Sigma_{D-2}$
\begin{equation}
    Q(\Sigma_{D-2}) = \int_{\Sigma_{D-2}} dS^{\mu\nu}   J_{\mu\nu} \label{eq:1formQ}
\end{equation}
and, as such, are always commuting. Hence, 1-form symmetries are always abelian, and are in correspondence with group elements of $U(1)$ when the symmetry is continuous. 

Local operators are not charged under 1-form symmetries, since they trivially commute with charges of the form \cref{eq:1formQ}. Line operators, however, can be charged under a 1-form symmetry $U(1)^{(1)}$\footnote{Henceforth, we use a superscript ${}^{(1)}$ to denote 1-form symmetry.}. Consider a line operator, $W_Q(C)$, with charge  $Q$ supported on a curve $C$, then the corresponding Ward identity is
\begin{align}
    \partial^\mu J_{\mu\nu}(x) W_Q(C) = Q \, \delta_{\nu}^{(3)}(x-C) W_Q(C). \label{eq:wilson-line-ward-identity}
\end{align}
where we have defined a higher-codimension delta function, 
\begin{align}
    \delta_{\nu}^{(3)}(x-C) = \int ds\, \frac{dy_\nu}{ds}\delta^{(4)}(x-y(s)) \,,
\end{align}
with $y(s)$ a given parameterization of the curve $C$, such that its integral over any cycle intersecting the curve equals one. 
    This is analogous to the contact term in the Ward identity of \cref{eq:nlsm-operator-eqns-general-operator} for local operators charged under a 0-form symmetry.
The Ward identity of \cref{eq:wilson-line-ward-identity} constitutes a non-trivial constraint on correlation functions involving the line operators $W_Q(C)$. We will show momentarily that it also implies the familiar soft photon theorem for scattering amplitudes in quantum electrodynamics.

1-form symmetries can be spontaneously broken if a charged operator develops a vacuum expectation value \cite{Gaiotto:2014kfa}
\begin{equation}
    \langle W_Q(C)\rangle \sim 1\,. \label{eq:vevW}
\end{equation}
This is equivalent to the perhaps more familiar ``perimeter law'' for a Wilson loop $\langle W_Q(C)\rangle \sim e^{-a L(C)}$, where $L(C)$ is the length of the loop, by a scheme choice in which a local counterterm on the line is added and tuned to cancel the perimeter scaling. \Cref{eq:vevW} indeed indicates a deconfined phase, and a version of Golstone's theorem \cite{Gaiotto:2014kfa,Lake:2018dqm} implies that the current must overlap with a one-photon state, which for helicity $h$ and momentum $p$ yields\footnote{The polarization vector appears complex conjugate because the current creates an incoming photon.} 
\begin{align}
    \langle \gamma_h(p)|J^{\mu \nu}(x)\rangle=\frac{i}{\emCoupling} \left( p^\mu\e_h^{\nu *}  -p^\nu\e_h^{\mu*}  \right)e^{ip \cdot x}\,, \label{eq:photoninterpol}
\end{align}
where $\emCoupling$ is the gauge coupling and $\e^\mu_h$ is the corresponding polarization vector, which satisfies the on-shell and normalization conditions
\begin{equation}
    \e\cdot q = 0\,, \qquad \e\cdot \e=0 \,, \qquad \e\cdot \e^*=-1\,.
    \label{eq:onshellpol}
\end{equation}
This is in precise correspondence with \Cref{eq:goldstone-interpolation}.

\Cref{eq:photoninterpol} implies that the current must contain a contribution linear in the NGB field,
\begin{align}
    J^{\mu\nu}(x) = \frac{1}{g^2} (\partial^{\mu}\photonField^\nu-\partial^{\nu}\photonField^\mu)\,+\cdots,
    \label{eq:JfromA}
\end{align}
where the dots denote possible higher-order contributions. 
Thus under $U(1)^{(1)}$ the photon transforms by a shift analogous to that of $\pi^a$ in \cref{eq:pion-shift}
\begin{align}
    \photonField_\mu(x) \mapsto \photonField_\mu(x) + \lambda_\mu(x)\, , \qquad \text{with} \qquad \partial_{[\mu}\lambda_{\nu]}(x) = 0\,.
    \label{eq:photon-shift}
\end{align}
Note that $\lambda_\mu(x)$ are the components of a closed, but not exact 1-form, so this is not equivalent to a small gauge transformation $A_\mu \to A_\mu + \partial_\mu \alpha$.

As in \cref{eq:chiral-field} the charge operators can be constructed by exponentiating the NGB field
\begin{align}
    W_Q(C) = \exp\left(i Q \int_C \photonField_\mu dx^\mu \right)\,,
\end{align}
so the NGB parameterizes fluctuations of the vacuum expectation value.
In the present case, these are of course the familiar expressions for the Wilson line operators, and by combining Eqs.~\eqref{eq:1formQ} and \eqref{eq:JfromA} their 1-form charges simply correspond to the electric flux.

Finally, we must discuss the fate of the 1-form symmetry in the presence of electrically charged matter.
    Indeed this induces the explicit breaking of the 1-form symmetry. The current for the 1-form symmetry is no longer conserved:
    \begin{equation}
       \partial^\mu J_{\mu\nu}(x) = j_\nu(x)\,, \label{eq:1formbreaking}
    \end{equation}
    where $j_\nu$ is the electromagnetic current.

Physically, the presence of electric charges allows the vacuum to polarize by pair production with the result that Wilson lines are screened and do not induce a non-trivial electric flux. Thus they cannot carry a 1-form charge.
This is closely related to the fact that the Wilson lines can now have endpoints on local charged operators, so the putative 1-form symmetry charges would act trivially on them.

\subsection{Soft photon theorems}

The famous soft photon theorem \cite{Low:1958sn,Weinberg:1964ew,Weinberg:1965nx,Burnett:1967km} concerns scattering amplitudes with charged particles. It states that the leading term in the soft expansion of an amplitude in the photon's momentum, $q$, is given by
\begin{align}
    \lim_{q\to0} \cA_{n+\gamma(q)} = \emCoupling \sum_i Q_i \frac{\e\cdot p_i}{q\cdot p_i}  \cA_{n} \label{eq:softphoton}
\end{align}
where $Q_i$ and $p_i$ are the charges and momenta of the hard particles.

In light of our discussion of breaking of 1-form symmetry by charged matter, it might naively seem that it is not possible to interpret \Cref{eq:softphoton} using this language. We are then forced to ask: is there any limit in which the 1-form symmetry re-emerges and explains the soft theorem? We will explain below that the answer to this question is positive, but let us begin by making some general comments that might give us hope that this is the case.

From the modern viewpoint, the masslessness of the photon is a consequence of the spontaneously broken 1-form symmetry and not of gauge invariance. From this perspective, one might also wonder why the photon remains massless even in the presence of charged particles. The answer is that 1-form symmetries are \emph{robust} in the sense that they cannot be broken by adding any local operators to the effective action, as these are not charged under the symmetry. Incidentally, this also implies that when 1-form symmetries emerge in some limit they are exact. This robustness protects the masslessness of the photon (though not completely, as it could be Higgsed by the charged matter).
More generally, the breaking of the 1-form symmetry is universal: it can only happen by adding charged degrees of freedom and in the form of \Cref{eq:1formbreaking}. One might expect that this universality is the moral reason for the existence of the soft photon theorems. In the rest of this section we will try to sharpen this perspective.

\subsubsection*{A photon Adler zero}

Let us first consider a trivial limit in which the 1-form symmetry is emergent. If we consider charged particles with mass $m$, then the scattering of photons with momenta $k\ll m$ is described by an effective theory \emph{\`a la} Euler-Heisenberg \cite{Heisenberg:1936nmg}, in which the charged matter is integrated out. This results in the non-linear Lagrangian
\begin{align}
    \mathcal{L}= -\frac{1}{4\emCoupling^2}F^{\mu\nu}F_{\mu\nu} + \frac{c_1}{m^4} (F_{\mu\nu}F^{\mu\nu})^2 +  \frac{c_2}{m^4} (F_{\mu\nu}F^{\nu\sigma}F_{\sigma\rho}F^{\rho\mu}) + {\cal O}\left(\frac{1}{m^6}\right)\,.  \label{eq:maxwell-lagrangian}
\end{align}
where $F_{\mu\nu} = \partial_\mu A_\nu - \partial_\nu A_\mu$ is the usual field strength, and $c_i$ are dimensionless Wilson coefficients which depend on the details of the matter that is integrated out.
This theory has an exact electric 1-form symmetry $U(1)_e^{(1)}$ with current 
\begin{align}
    J_e^{\mu\nu} = \frac{1}{\emCoupling^2}F^{\mu\nu} + {\cal O}(F^3), \label{eq:maxwell-higher-form-current-e}
\end{align}
conserved by the equations of motion derived from \eqref{eq:maxwell-lagrangian}. It also has a magnetic $U(1)_m^{(1)}$ 1-form global symmetry with the current
\begin{align}
    J_m^{\mu\nu} = \frac{1}{4\pi}\epsilon^{\mu\nu\alpha\beta}F_{\alpha\beta} \,,\label{eq:maxwell-higher-form-current-m}
\end{align}
conserved by the Bianchi identity and which measures magnetic flux. The NGB associated to $U(1)_m^{(1)}$ is a dual photon which does not appear explicitly in \cref{eq:maxwell-lagrangian} but likewise transforms non-linearly by a shift under $U(1)_m^{(1)}$. This symmetry will play an important role in \cref{sec:new-soft-theorems-from-2-group}, but not in our current discussion.

Let us now follow the steps analogous to the derivation of the pion Adler zero. We separate the single-photon and hard contributions in the form factor for the 1-form symmetry current  within a multi-photon state $|\alpha \rangle$,
\begin{align}
    \langle \alpha | J_e^{\mu\nu}(q)\rangle=&\frac{1}{g} \frac{ (q^\mu \e^{\nu*} - q^\nu \e^{\mu*})}{q^2}\langle \alpha+ \gamma(q)|0 \rangle+\langle \alpha| J^{\mu\nu}_{e H}(q)\rangle\,. \label{eq:1-form-current-decomposition}
\end{align}
On the RHS we have left a sum over helicities implicit. The coefficient of the pole is simply related to the scattering amplitude with an additional photon with momentum $q$
\begin{equation}
    \langle \alpha+ \gamma(q)|0 \rangle  = i \cA_{n+\gamma(q)}\,.
\end{equation}
Then the conservation of the 1-form current implies the relation
\begin{align}
 \cA_{n+\gamma(q)} = -i g \,q_\mu \e_\nu \langle \alpha| J^{\mu\nu}_{e H}(q)\rangle\, ,
\end{align}
where we made use of \Cref{eq:onshellpol}. Taking the soft limit we find that the RHS vanishes, since as before, the hard current contributions are regular in the soft limit due to the shift symmetry of the NGB implying that there is no cubic coupling. 
Thus we find an Adler zero for photon amplitudes in this theory
\begin{align}
    \lim_{q\to0} \cA_{n+\gamma(q)} = 0 
\end{align}
This is in exact analogy with Eqs.~\eqref{eq:axial-current-conservation} and~\eqref{eq:adler-zero-nlsm} in the previous section. 

Such a zero is perhaps an unsurprising statement, since the effective theory in \cref{eq:maxwell-lagrangian} is derivatively coupled with at least one derivative by field, that is, it is a theory of abelian NGB. This soft photon theorem is, nevertheless, a direct consequence of the electric 1-form symmetry of the theory.

\subsubsection*{Low's soft photon theorem}
By integrating out the charged matter we have seemingly thrown away the baby with the bathwater, and have only been able to reproduce \Cref{eq:softphoton} in the case where the scattering amplitude involves no external charged particles. However there is a middle way in which we can keep both external charged states and see the 1-form symmetry emerge.

Consider a scattering amplitude involving $n$ massive charged particles with `hard' momenta, $p_i$. In the limit that all hard momenta are neighboring the mass shell of either external particle or anti-particle states, we can write
$
    \ell_i^\mu= p_i^\mu + k^\mu
$
with $p_i$ on-shell and $k_i\ll p_i$. Expanding in this limit, the scattering amplitude will be  given by a correlation function of Wilson lines plus a hard (local) operator insertion, $O_H$,
\begin{align}
    &i\cA_{n} \sim \langle W_{Q_1}(C_1)\dotsm W_{Q_n}(C_n) O_H\rangle 
    \label{eq:wilson-lines-in-correlator}
\end{align}
where $\sim$ denotes that the equality holds up to corrections in inverse powers of the masses.
The Wilson lines are extended from the origin to infinity along trajectories with four-velocities $v_i^\mu = p_i^\mu/m_i$ (that is, $C_i: y_i^\mu(s) = s v^\mu_i$) with $s\in[0,\infty]$, and carry the corresponding charges, $Q_i$,
\begin{equation}
    W_{Q_i}(C_i) = \text{exp}\left(iQ_i\int_{0}^{\infty} ds\, v_i^\mu A_\mu(sv)\right)
\end{equation}
The hard operator $O_H$ is inserted at the origin. Its form will not be important in what follows.

We claim that the electric 1-form symmetry emerges in this limit. We will show that this is the case momentarily, but first let us consider the consequence of the Ward identity \cref{eq:wilson-line-ward-identity} upon the correlation function appearing in the expanded amplitude \cref{eq:wilson-lines-in-correlator}
\begin{align}
\partial_\mu \langle J_e^{\mu\nu}(x) \prod_i W_{Q_i}(C_i) O_H\rangle =  \left(\sum_i Q_i\int ds\, \frac{dy_i^\nu}{ds}\delta^{(4)}(x-y_i(s))\right) \langle \prod_i W_{Q_i}(C_i) O_H\rangle\,. \label{eq:fourier-transform-one-wilson-line}
\end{align}
We will contract this equation with a polarization $\e^\nu(q)$, and Fourier transform in $x$. On the one hand, the LHS yields,
\begin{align}
    &\e_\nu \int d^4x \, e^{-iq\cdot x} \partial_\mu\langle  J_e^{\mu\nu}  \prod_i W_{Q_i}(C_i) O_H\rangle \\
    &\hspace{2cm}= -\frac{i}{\emCoupling}\langle \gamma_h(q)| \prod_i W_{Q_i}(C_i) O_H\rangle + i q_\mu \e_\nu \langle J^{\mu\nu}_{eH}(q)  \prod_i W_{Q_i}(C_i) O_H\rangle \, , \label{eq:1-form-ward-identity-lhs}
\end{align} 
where the the derivative operator has effected an amputation of an external photon $\gamma_h(q)$.
The first term is then identified with the amplitude with an additional external photon
\begin{equation}
   \langle \gamma(q)| \prod_iW_{Q_i}(C_i) O_H\rangle \sim i  \cA_{n+\gamma(q)} 
\end{equation}
On the RHS of \cref{eq:fourier-transform-one-wilson-line} we just need to compute the Fourier transform of the contact terms
\begin{align}
    \int d^4x \, e^{-iq\cdot x} \left(Q_i\int_{0}^{\infty} ds\,\e\cdot v_i \delta^{(4)}(x - v_is)\right)  =  Q_i \e\cdot v_i \int_{0}^{\infty} ds\, e^{-is q\cdot v_i} = -i Q_i \frac{\e\cdot v_i}{q\cdot v_i}  \label{eq:wilson-line-ward-fourier-transform}
\end{align}
which shows that the the RHS of the Ward identity gives
\begin{align}
   -\sum_i iQ_i\frac{\e\cdot v_i}{q\cdot v_i} \, \langle \prod_i W_{Q_i}(C_i) O_H\rangle & \sim  \sum_i Q_i\frac{\e\cdot p_i}{q\cdot p_i} \cA_{n}
\end{align}
Putting it all together yields the relation
\begin{align}
    \cA_{n+\gamma(q)}  = \emCoupling \sum_i Q_i\frac{\e\cdot p_i}{q\cdot p_i}  \cA_{n} - i \emCoupling q_\mu \e_\nu \langle J^{\mu\nu}_{eH}(q)  \prod_i W_{Q_i}(C_i) O_H\rangle\,.
\end{align}
In the soft limit the matrix element of the hard current can only give terms which are at most singular as ${\cal O}(q^{-1})$, so taking the small $q$ limit of this relation gives the soft theorem in \Cref{eq:softphoton}.
\subsection{Emergent higher-form symmetry in heavy-particle EFT}

Above we have claimed that in the limit where the soft photon leaves particles (or antiparticles) close to the mass shell there is an emergent 1-form symmetry, which we then used to derive the soft photon theorem. Let us now make this precise using the language of effective field theory. The manipulations in this section are not new, and very familiar in the context of HQET \cite{Manohar:2000dt}, but we will describe them emphasizing the physics of the emergence of the 1-form symmetry.

\begin{figure}[t]
    \centering
         \begin{tikzpicture}[scale=0.7]
         \draw[line width=1.3cm, color=teal!30]
         plot[domain=-5:5,
         samples = 50,
         smooth]({\x}, {1.5+0.07*\x*\x});
         \draw[ultra thick,teal]
         plot[domain=-5:5,
         samples = 50,
         smooth]({\x}, {1.5+0.07*\x*\x});
         \draw[ultra thick,teal]
         plot[domain=-5:5,
         samples = 50,
         smooth]({\x}, {-1.5-0.07*\x*\x});
         \draw[very thick,-stealth] (0,-4)-- (0,4) node[right]{$\ell^0$};
         
         \draw[very thick,-stealth] (-5,0) -- (5,0) node[below]{$|\vec \ell|$};

         \draw[thick,-latex] (0,0) -- (3,2.15) ;
         \draw (1.2,0.8) node[above]{$p^\mu$};
         \draw[thick,-latex] (3,2.15) -- (3.4,1.5) ;
         \draw  (3.2,2) node[right]{$k^\mu$};
         \draw[thick,-latex] (0,0) -- (3.4,1.5) ;
         \draw (1.9,0.85) node[below]{$\ell^\mu$};

        \draw (-4.25,-2.6) node[left]{$\varphi_v^-$};
        \draw (-4.25,2.6) node[left]{$\varphi_v^+$};
         
         \draw (-2.4,2.9) node[left,rotate=-24,anchor=north]{$p^2=m^2$};
        
        \end{tikzpicture}
        \caption{Illustration of the decomposition of a momentum $\ell^\mu=p^\mu+k^\mu$, as a large component on the mass shell, $p^\mu$, and a small component $k^\mu \ll p^\mu$. The field $\varphi_v^+$ describes small fluctuations around the mass shell in the shaded region, whereas all modes outside this region are integrated out in the effective theory.}
    
    \label{fig:heft}
\end{figure}
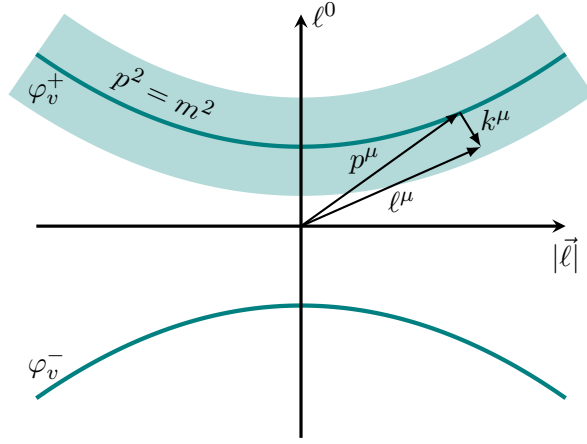

We begin by considering a charged massive particle associated to a field $\phi(x)$ interacting with the photon. In order to manifest the excitations close to the mass shell we perform a field redefinition which pulls out the large part of the momentum 
\begin{equation}
    \phi(x) = \frac{e^{i m v \cdot x}}{\sqrt{2m}} \,(\varphi^+_v + \varphi^-_{v})
\end{equation}
where $v^\mu = p^\mu/m$ with $p$ on-shell, and we have also separated the field into positive and negative frequency modes corresponding to the particle and antiparticle.\footnote{These can be extracted using projection operators $\varphi^\pm_v= P_\pm\phi$ which depend on the spin of the particle. For instance, for scalars $P_\pm \propto (i v\cdot D \pm m)$ and for Dirac fermions $P_\pm \propto 1 \pm v_\mu \gamma^\mu$.} This field redefinition turns large derivatives of the field $\partial \phi \sim p \phi \sim m \phi $ into manifest couplings, and leaves only small derivatives $\partial \varphi \sim k \varphi \ll m \varphi$ in the effective Lagrangian
\begin{equation}
    {\cal L}= -\frac{1}{4\emCoupling^2}F^{\mu\nu}F_{\mu\nu} + i\varphi^{+*}_v (v\cdot D)\varphi^+_v +  i\varphi^{-*}_v (v\cdot D+2m)\varphi^-_v + {\cal O}\Big(\frac{1}{m}\Big)\,,
\end{equation}
where $D_\mu = \partial_\mu + i A_\mu$ is the usual covariant derivative.
Thus we see that the positive frequency $\varphi^+_v$ fields describe excitations close to the particle mass shell, and $\varphi^-_v$ describe antiparticle excitations which have a large gap $k \sim 2m$. This is illustrated in Fig.~\ref{fig:heft}. We can integrate out the antiparticle excitations to get an effective theory describing only particles
\begin{equation}
    {\cal L} = i\varphi_v^* (v\cdot D)\varphi_v + \varphi_v^* \frac{D_\perp^2}{2m}\varphi_v +  d\,  \varphi_v^* \frac{F_{\mu\nu} M^{\mu\nu}}{4m}\varphi_v + \cdots\,.
\end{equation}
Here we have kept terms only up to subleading power in $1/m$, including the kinetic energy written in terms of $D_\perp^\mu = D^\mu - v^\mu (v\cdot D)$; and the magnetic dipole moment operator with Wilson coefficient $d$, written using the Lorentz generator, $M^{\mu\nu}$, in the appropriate representation.

At this point an astute reader might guess that this theory has an emergent 1-form symmetry. As explained above, the explicit breaking of  1-form symmetries by charged matter is closely related to the phenomenon of pair creation or vacuum polarization. However in constructing this effective theory we have integrated out the antiparticles, so such processes cannot occur. A further consequence of this is that the coupling constant $g$ does not run in the EFT and is frozen at the matching scale, $m$, which makes the current $J^{\mu\nu} = \frac{1}{\emCoupling^2} F^{\mu\nu} + \cdots$ well-defined.

The symmetry can be made manifest by performing one further field redefinition (due to BPS \cite{Bauer:2001yt}) which strips a Wilson line from the particle fields
\begin{equation}
    \varphi_v = W_Q(C)  \tilde \varphi_v =   \text{exp} \left(iQ \int_{0}^\infty v^\mu A_\mu(s v)\right) \tilde \varphi_v\,.
\end{equation}
This justifies our claim in \Cref{eq:wilson-lines-in-correlator} that the amplitude of charged particles can be related to a correlation function of Wilson lines. Furthermore, by the identity\footnote{The notation with $1/(v\cdot \partial)$ is shorthand for operator insertions on the Wilson line, e.g.,
\begin{equation*}
    \frac{1}{v\cdot \partial} F_{\mu\nu}(x)= \int^{\infty}_0 ds\,  F_{\mu\nu}(x+ s v) \,.
\end{equation*}
}
\begin{equation}
    D_\mu \varphi_v = W_Q(C) \left(\partial_\mu +\frac{1}{v\cdot\partial} F_{\mu\nu}v^\nu \right) \tilde \varphi_v\,, 
\end{equation}
it removes the dependence on bare gauge fields without derivatives in the effective Lagrangian
\begin{equation}
    {\cal L} = \tilde\varphi_v^* (v\cdot \partial) \tilde\varphi_v + \tilde \varphi_v^* \frac{1}{2m} \left(\partial^\perp_\mu +\frac{1}{v\cdot\partial} F_{\mu\nu}v^\nu \right)^2 \tilde\varphi_v +  d\,  \tilde{\varphi}_v^* \frac{F_{\mu\nu} M^{\mu\nu}}{4m}\tilde{\varphi}_v  + \cdots\,,
\end{equation}
which now manifestly exhibits the shift symmetry in the gauge field of \Cref{eq:photon-shift}, as it only depends on field strengths. In fact, this observation is not restricted to the leading powers in the $k/m\ll 1$ limit, but holds to all orders. We have shown that, as expected, the one-form symmetry emerges as an exact symmetry in the heavy-particle limit, which is dual to the soft photon limit, justifying our analysis of soft photon theorems.

While we have focused here on the case of emergent $U(1)$ 1-form symmetries in theories with an abelian gauge field, we expect this to be a generic feature of theories in which anti-particle production is suppressed and Wilson lines play a significant role. For instance, we think it is likely that the discrete 1-form symmetries of pure Yang-Mills theory emerge as symmetries of HQET or of the ultrasoft sector of Soft Collinear Effective Theory (SCET) \cite{Bauer:2000yr}.

\label{sec:higher-form-lorentzian-eft}

\section{New soft theorem from higher-group symmetry}
\label{sec:new-soft-theorems-from-2-group}
The fact that higher-form symmetries are always abelian means that the operator equations satisfied by their currents do not contain non-trivial contributions at coincident points,\footnote{'t Hooft anomalies are, of course, the exception to this rule.} and hence there are no interesting double soft theorems for the associated NGB. However there exists a generalized symmetry structure known as higher-group global symmetry which mixes global symmetries of different rank in a non-trivial way \cite{Kapustin:2013uxa,Baez2004,Cordova:2018cvg,Nakajima:2022jxg,Antinucci:2024bcm,Davighi:2024zjp}. We will focus on the case of 2-group symmetry and show that, when spontaneously broken, such structure leads to interesting double soft theorems.

The hallmark of a continuous 2-group symmetry is an additional contact term in the Ward identity of 0-form currents which depends on the 1-form symmetry current
\begin{align}
  \partial^\mu  j_\mu^a(x)j^b_\nu(y) \supset i\frac{\kappa}{2\pi}\delta^{ab}\partial^\lambda\delta^{(4)}(x-y)J_{\nu\lambda}(y), \label{eq:2-group-new-term}
\end{align}
where $\kappa$ is a quantized constant. The 2-group symmetry is denoted by $G\times_\kappa U(1)^{(1)}$, where $G$ is generated by the $j^a$ and $U(1)^{(1)}$ is a 1-form symmetry generated by $J$. Due to the form of the Ward identity the 0-form currents do not form a closed subalgebra, so the group $G$ does not label a 0-form symmetry. One might think of the 2-group as a non-trivial extension of the 0-form $G$ by the 1-form $U(1)^{(1)}$.\footnote{The 0-form analog of this is the fact that a coset $G/H$ need not be a subgroup of $G$, but one can think of $G$ as an extension of $G/H$ by $H$.}

Intuitively, one might think of $\kappa$ as an $f^{abc}$ structure constant in which one of the indices points in the 1-form ``direction''. Hence, when the symmetry is spontaneously broken one can expect that the double soft theorem will describe mixing of the 0-form NGB (pions) and the 1-form NGB (photons). Furthermore, the fact that the contact term in \Cref{eq:2-group-new-term} has an additional derivative suggests that the 2-group structure will be visible at the sub-leading order. We will see that this is precisely correct.

\subsection{2-group double soft pion theorem}

In what follows we  focus on a theory possessing a 2-group global symmetry,
\begin{align}
    \left(G_L\times G_R\right)\times_\kappa U(1)_m^{(1)}\, . \label{eq:2-group-symmetry}
\end{align}
where  $G_L\times G_R$ is a non-abelian chiral group, and $U(1)_m^{(1)}$ a magnetic 1-form symmetry.  When this 2-group symmetry structure is spontaneously broken to $G_V$, the low-energy dynamics are universally described at low energies by a theory of NGB featuring pions  $\pi^a$ associated to $G_L\times G_R$, as well as a photon associated to the $U(1)^{(1)}_m$, and currents
    \begin{align}
        \jA^a_\mu(x) = f_\pi \partial_\mu\pi^a(x) + \cdots\, \qquad \textrm{and}\qquad
        J_{\mu\nu}(x)= \frac{1}{4\pi}\epsilon_{\mu\nu\alpha\beta}F^{\alpha\beta}(x)\,,
    \end{align}
as well as a vector current $V_\mu^a(x)$ quadratic in the fields.
The 2-group is encoded in the structure of the Ward identities
\begin{align}
    \partial^\mu  V_\mu^a(x)V^b_\nu(y) &= if_T^{abc}\delta^{(4)}(x-y)  V^c_\nu(y)\\
    \partial^\mu  \jA_\mu^a(x)\jA^b_\nu(y) &= iF^{abc}\delta^{(4)}(x-y)  V^c_\nu(y)\\
    \partial^\mu  \jA_\mu^a(x)V^b_\nu(y) &= if_X^{abc}\delta^{(4)}(x-y) \jA^c_\nu(y)-i\frac{\kappa}{2\pi}\delta^{ab}\partial^\lambda\delta^{(4)}(x-y)J_{\nu\lambda}(y)\,. \label{eq:ward-identity-2-group-V-A}
\end{align}
The new term appearing in \cref{eq:ward-identity-2-group-V-A} contains the symmetry current associated to magnetic 1-form symmetry $U(1)_m^{(1)}$ and the quantized constant $\kappa$. 

The effective Lagrangian realizing such symmetry breaking pattern\cite{Antinucci:2024bcm} is,
\begin{align}
    \cL = \,&\frac{f_\pi^2}{4}\tr\left[\partial_\mu U \partial^\mu U^{\dagger}\right] - \frac{1}{4\emCoupling^2} F_{\mu\nu}F^{\mu\nu}
    \label{eq:lagrangian-2-group}
    \\
    & \hspace{1cm}-\frac{i\kappa}{24\pi^2} \,\epsilon^{\mu\nu\alpha\beta}\photonField_\mu\tr\left[(iU^{\dagger}\partial_\nu U)(iU^{\dagger}\partial_\alpha U)(iU^{\dagger}\partial_\beta U)\right] +\cdots\,,\nn
\end{align}
Comparing to the action for 0-form spontaneous breaking \cref{eq:lagrangian-nlsm}, we have added a dynamical $U(1)$ gauge field -- a photon -- and a term which couples it to the pion. One consequence of this coupling is to explicitly break the putative 1-form $U(1)_e^{(1)}$ symmetry associated to the photon. However $U(1)_m^{(1)}$ is the 1-form symmetry participating in the 2-group \cref{eq:2-group-symmetry}. This means that strictly speaking the photon is a pseudo-NGB and the dual photon is a NGB.

The pions are uncharged under the $U(1)$ gauge group, however the pion-photon interaction term in \cref{eq:lagrangian-2-group} is a coupling of the photon to a topological symmetry current,
\begin{equation}
   B^\mu = i\frac{\kappa}{24\pi^2} \,\epsilon^{\mu\nu\alpha\beta}\tr\left[(iU^{\dagger}\partial_\nu U)(iU^{\dagger}\partial_\alpha U)(iU^{\dagger}\partial_\beta U)\right]. 
   \label{eq:top-current-B}
\end{equation}
In the absence of a dynamical photon, and hence in the absence of the 2-group global symmetry, the current $B^\mu$ would generate a topological 0-form $U(1)$ global symmetry. 

This theory is far from exotic, as for $G_L\times G_R=SU(N_f)_L\times SU(N_f)_R$ it describes the low-energy limit of massless QCD with gauged $U(1)_V$ vector symmetry corresponding to Baryon number, which acts diagonally on quarks.\footnote{Technically, one must also add a Wess-Zumino-Witten term to match the various 't Hooft anomalies in the chiral symmetry, but we drop this here for simplicity.}
Such symmetry, with current \eqref{eq:top-current-B}, is associated to the non-trivial homotopy group $\pi_3(SU(N_f))=\mathbb{Z}$, and the field configurations where $B^\mu$ integrates to a non-trivial winding number are identified with baryons \cite{Witten:1983tx}. In fact, the photon-pion coupling term was first suggested in \cite{Witten:1983tw}, where it was introduced as an anomalous contribution to the baryon current in the chiral effective theory.\footnote{Note that our photon is not the usual photon of electromagnetism, which arises from gauging a linear combination of $U(1)_V$ and a $U(1)$ subgroup of $SU(N_f)_V$. An easy way to see this is that the pions in our effective theory are not charged under $U(1)_V$. As a consequence, the $U(1)_V$ photon does not mediate a long-range interaction between the pions.} As explained in \cite{Cordova:2018cvg}, $2$-group global symmetries may arise from gauging a $0$-form global symmetry with a mixed anomaly. This is possible, for example, when the original theory possesses a mixed 't Hooft anomaly which is quadratic in a non-abelian $G$ and linear in a $U(1)$ symmetry, which is then gauged. Indeed this is the case in massless QCD where there is a mixed anomaly between the $U(1)_V$ and $SU(N_f)_L\times SU(N_f)_R$ which gives rise to the 2-group structure.

Let us comment on the discrete symmetries of this theory. In the absence of the pion-photon interaction term the theory has four distinct  $\mathbb{Z}_2$ symmetries: parity, $P_0:x_i \mapsto -x_i$ for $i = 1,2$, and 3, charge conjugation of the photon ${C_1:\photonField_\mu \mapsto -\photonField_\mu}$, charge conjugation of the chiral field ${C_2:U\mapsto U^T}$, and pion number mod\nobreakdash-2 ${(-1)^{N_\pi}:U\mapsto U^{-1}}$ (or $\pi^a\to-\pi^a$). The pion-photon interaction breaks one of these, leaving the $(\mathbb{Z}_2)^3$ discrete symmetry associated to the combinations \cite{Antinucci:2024bcm}
\begin{align}
    P = P_0(-1)^{n_\pi}, \qquad C =C_1C_2, \qquad \widetilde{C}=C_1(-1)^{n_\pi}\,,
    \label{eq: discrete_symmetries}
\end{align}
which include a new parity, $P$, and charge conjugation $C$, as well as pion $+$ photon number mod-2, $\widetilde C$.
We see that the 2-group theory allows, for example, scattering amplitudes involving an odd-number of pions so long as there is also an odd-number of photons such that $\widetilde{C}$ is conserved.

The main result of this paper is a new double soft pion theorem for amplitudes with $n_\pi+2$ pions and $n_\gamma$ photons stemming from the continuous 2-group symmetry, which takes the form
\begin{align}
   \lim_{q_1,q_2\rightarrow0}\mathcal{A}_{(n_\pi+\pi^a( q_1)\pi^b( q_2),n_\gamma)} = \left(S^{(0)} + S^{(1)}+S^{(1)}_{\kappa}\right) \mathcal{A}_{(n_\pi,n_\gamma)} \label{eq:double-soft-theorem-2-group-schematic}
\end{align}
where the $S^{(i)}$ are those given in \cref{eq:NLSM-double-soft-subleading-order} and
\begin{align}
S^{(1)}_{\kappa}\mathcal{A}_{(n_\pi,n_\gamma)}=
& \frac{i\kappa}{2f_\pi^3\pi^2}\sum^{n_\pi}_{i=1}\sum_{\textrm{h}}f^{ab a_i}\frac{\epsilon( q_1 q_2 p_i \e^*_{ih})}{2p_i\cdot(q_1+q_2)}\mathcal{A}^{a_1\ldots a_{i-1}a_{i+1}\ldots a_n}_{(n_\pi-1,n_\gamma+1)} \nonumber \\
& -\frac{i\kappa}{2f_\pi^3\pi^2}\sum^{n_\gamma}_{j=1}f^{ab d}\frac{ \epsilon( q_1q_2 k_j \e_j)}{2k_j\cdot(q_1+q_2)}\mathcal{A}^{da_1\ldots a_n}_{(n_\pi+1,n_\gamma-1)}\,,
\label{eq:double-soft-next-to-leading-order-2-group}
\end{align}
where in the first line we sum over helicities of internal photons and we introduced the notation $\epsilon(abcd)=\epsilon^{\mu\nu\rho\sigma}a_\mu b_\nu c_\rho d_\sigma$. This new soft factor $S^{(1)}_\kappa$ is the consequence of the spontaneously broken 2-group global symmetry. $S^{(1)}_\kappa$ is odd under the parity $P_0$ of the theory without the pion-photon interaction. However the full RHS of \cref{eq:double-soft-next-to-leading-order-2-group} is invariant under $P=P_0(-1)^{n_\pi}$ due to the lower-point amplitudes, $\mathcal{A}_{n_\pi-1,n_\gamma+1}$ and $\mathcal{A}_{n_\pi+1,n_\gamma-1}$, having one fewer and one more pion, respectively. Acting on the lower-point amplitude, $S^{(1)}_\kappa$ effects a `rotation' of pions into photons, and vice versa. As we will see, the presence of those terms is required precisely by the appearance of the new term with the 1-form symmetry current in the 2-group Ward identities in \cref{eq:ward-identity-2-group-V-A}. We also note that the non-trivial 2-group soft factor depends on the antisymmetric non-abelian structure constant of the 0-form group, and thus at this order we do not expect an analogous non-trivial 2-group soft factor for abelian 2-groups \cite{Brennan:2023mmt}.

\subsection{Proof}
We will now explain how the derivation of the double soft theorem is augmented by the 2-group. We emphasize that since the proof uses current algebra it is valid to all loop orders.\footnote{The sub-leading soft factor $S^{(1)}_\kappa$, from the 2-group modified current algebra, is valid to all loop orders. However the soft factor $S^{(1)}$ from the NLSM current algebra is corrected at 1-loop, as we discuss in \Cref{app:soft-axial-nlsm}.} As in \cref{sec:0-form}, our starting point is the following consequence of the Ward identity,
\begin{align}
    q_1^\mu q_2^\nu\langle\alpha| \jA^a_\mu(q_1)\jA^b_\mu(q_2) \rangle=-\frac{1}{2}F^{abc}(q_1-q_2)^\mu\langle \alpha|V^c_\mu(q_1+q_2) \rangle. \label{eq:ward-two-axial-currents-2-group}
\end{align}
Notice that this relation is unmodified by the 2-group current algebra, since the Ward identity involving two axial currents is unmodified. Thus, the intermediate relation \cref{eq:double-soft-correlators} follows in the new theory,
\begin{align*}
    &f^2_\pi \langle\alpha + \pi^a(q_1)\pi^b(q_2)|0\rangle = \frac{1}{2}F^{abc}(q_1-q_2)^\mu\langle \alpha| V^c_\mu(q_1+q_2)\rangle + q_1^\mu q_2^\nu\langle \alpha| \jAbar^a_\mu(q_1)\jAbar^b_\nu(q_2)\rangle\, . \label{eq:double-soft-correlators-in-2-group-section}
\end{align*}
Next, we need to analyze additional contributions to $\langle \alpha| V^c_\mu(q)\rangle$ and $\langle \alpha| \jAbar^a_\mu(q_1)\jAbar^b_\nu(q_2)\rangle$ in the new theory.

\subsubsection*{Soft vector form factor}
We follow the same strategy as in the case of spontaneous 0-form symmetry breaking (see Appendix \ref{app:soft-vector-current}) and insert a complete set of states into the form factor, where $\langle \alpha|$ now denotes a state with $n$ pions and $m$ photons. The dominant contribution in the soft limit comes from single-particle states. We have
\begin{align}
   \lim_{q\to0} \langle \alpha|V^a_{\mu}(q)\rangle
=&\sum^{n+m}_{i=1}\sumint_{X}\langle X_i|V^a_\mu(q)|X\rangle\Delta_X\langle X+\hat\alpha_i|0\rangle\,,
\end{align}
where by $\hat \alpha_i$ we denote the state $\alpha$ with the $i$-th particle $X_i$ removed, $|X\rangle$ stands for a single-particle state of either pions or photons, and $\Delta_X$ is the corresponding propagator.

The $\langle\pi^b(p)|V^a_\mu(q)|\pi^c(k)\rangle$ form factor is the same as in the 0-form case derived in Appendix \ref{app:soft-vector-current}. Since there is no flavor-structure corresponding to $\langle\gamma_h(p)|V^a_\mu(q)|\gamma_{h'}(k)\rangle$, the form factor vanishes. Thus the only additional form factor we need is $\langle \pi^{b}(p)| V^a_\mu(q) |\gamma_h(k)\rangle$, which at leading order in soft momentum can be parametrized as
\begin{align}
    \langle \pi^{b}(p)| V^a_\mu(q)|\gamma_h(k) \rangle =A(p,q)B^{ab}\epsilon_{\mu\nu\rho\sigma}q^\nu p^\rho \e_h^{*\sigma}+\mathcal O(q^2)\,, \label{eq:vector-pi-gamma-ansatz}
\end{align}
where $k=q+p$, $A(p,q)$ is a Lorentz-invariant structure function and $B^{ab}$ is an arbitrary flavor-structure. Here we have specialized to an ansatz that is odd under parity, $P_0$. As discussed above, the theory is invariant under the product of any two of the four $\mathbb{Z}_2$ transformations: $C_1, C_2, P_0$, and $(-1)^{N_\pi}$. It follows that the form factor \cref{eq:vector-pi-gamma-ansatz} must be $P_0$-odd. We can constrain the coefficient in the ansatz by considering a related object $\langle\jA_\nu^{b}(p) V^a_\mu(q) | \gamma_h(k)\rangle$ and its axial current decomposition.

We write an ansatz for $\langle \jA_\nu^{b}(p) V^a_\mu(q) |\gamma_h(k)\rangle$ and impose the Ward identity for the axial current
\begin{align}
    p^\nu\langle \jA_\nu^{b}(p) V^a_\mu(q) |\gamma_h(k)\rangle = -i\frac{\kappa}{2\pi} \delta^{ab} p^\nu\langle J_{\mu\nu}(p+q)|\gamma_h(k)\rangle\,,
\end{align}
where we used the fact that $\langle \jA_{\mu}(p+q)|\gamma_h(p+q)\rangle=0$. Similarly, we require that the Ward identity for the vector current is satisfied. Those two conditions allow us to constrain the correlator at the leading order in soft momentum
\begin{align}
    \langle  V^a_\mu(q) \jA_{\nu}^{b}(p)|\gamma_h(k)\rangle =&-\delta^{ab}\frac{\kappa}{8\pi^2f_\pi}\left(\epsilon_{\mu\nu\sigma\lambda}(p+q)^\lambda-\frac{2 p_\nu}{p^2}\epsilon_{\mu\lambda\rho\sigma}q^\lambda p^\rho\right)\e^{*\sigma}_h+\mathcal O{(q^2,p^2)}\,,
    \label{eq:2-group-vector-result}
\end{align}
which in turn implies, using \cref{eq:axial-current-decomposition}, that
\begin{align}
    \langle \pi^{b}(p)| V^a_\mu(q)|\gamma_h(k) \rangle =\delta^{ab}\frac{\kappa}{4\pi^2f_\pi}\epsilon_{\mu\nu\rho\sigma}q^\nu p^\rho \e^{*\sigma}_h\,+\cO(q^2)\,.
    \label{eq:Vpi_2-group_result}
\end{align}
Finally, putting everything together we arrive at the soft limit of the vector current in the theory with a 2-group global symmetry, 
\begin{align}
\lim_{q\rightarrow 0}\langle \alpha|V^a_{\mu}(q)\rangle=&-i\sum_{i=1}^{n_\pi} f^{a a_i d}\left(\frac{(2p_{i}+q)_\mu}{(p_i+ q)^2} -\frac{iq^\nu \angMom_{i\mu\nu}}{(p_i\cdot q)} \right)\cA_{(n_\pi,n_\gamma)}^{a_1\ldots d\ldots a_{n_\pi}}\nn\\
&-\frac{\kappa}{4\pi^2f_\pi}\sum_{i=1}^{n_\pi}\sum_{h}\delta^{aa_i}\frac{\epsilon_{\mu\nu\rho\sigma}q^\nu p_i^\rho \e^{*\sigma}_i }{2p_j\cdot q}\cA^{a_1\ldots a_{i-1}a_{i+1}\ldots a_{n_\pi}}_{(n_\pi-1,n_\gamma+1)}\nn\\
&+\frac{\kappa}{4\pi^2f_\pi}\sum_{j=1}^{n_\gamma} \delta^{aa_i}\frac{\epsilon_{\mu\nu\rho\sigma}q^\nu k_j^\rho \e_j^{\sigma}}{2 k_j\cdot q}\cA^{da_1\ldots a_{n_\pi}}_{(n_\pi+1,n_\gamma-1)}\,,
\label{eq:soft-vector-current-theorem-2-group}
\end{align}
where $\langle \alpha|$ again denotes a state with $n_\pi$ pions and $n_\gamma$ photons. The first line of \cref{eq:soft-vector-current-theorem-2-group} is the contribution from the $G_L\times G_R$ symmetry as derived in Appendix \ref{app:soft-vector-current}.

\subsubsection*{Soft axial-axial hard current form factor}
\label{sec:soft-axial-axial-2-group}
Repeating the same analysis for the soft axial-axial hard current form factor $\langle \alpha| \jAbar ^a_\mu(q_1)\jAbar ^b_\nu(q_2)\rangle$ we see that the new relevant form factor is $\langle \pi^{c}(p)|\jAbar ^a_\mu(q_1)\jAbar ^b_\nu(q_2)|\gamma_h(k)\rangle$,\footnote{The form factor $\langle \gamma_h |\jAbar^{a}_\mu \jAbar^b_\nu|\gamma_{h'}\rangle$ is suppressed in the soft limit, as can be deduced from an analogous bootstrap argument imposing Ward identities. One can also explicitly perform a perturbative check and see that the loop diagrams contributing to $\langle \gamma_h |\jAbar^{a}_\mu \jAbar^b_\nu|\gamma_{h'}\rangle$ are suppressed in the soft limit.} which can be obtained from the 2-group current algebra. As is consistent with the discrete $\mathbb{Z}_2$ symmetry structure of the theory, we start with the $P_0$-odd ansatz
\begin{align}
    \langle \pi^c(p) | \jAbar_\mu^a(q_1)\jAbar_\nu^b(q_2)|\gamma_h(k) \rangle = A(p,q_1,q_2)B^{abc}\epsilon_{\mu\nu\rho\sigma}p^\rho \e^{*\sigma}_h+\mathcal{O}(q_1,q_2)\,,
\end{align}
where $A(p,q_1,q_2)$ is a Lorentz-invariant structure function and $B^{abc}$ is an arbitrary flavor-structure.
We consider a related object,
\begin{align*}
    \langle  \jA^c_\lambda(p)\jA_\mu^a(q_1)\jA_\nu^b(q_2)|\gamma_h(k)\rangle\,,
\end{align*}
which we constrain using the Ward identity
\begin{align}
    p^\lambda\langle \jA^c_\lambda(p)\jA_\mu^a(q_1)\jA_\nu^b(q_2)|\gamma_h(k)\rangle=&F^{cae}\langle V^e_\mu(p+q_1)\jA_\nu^b(q_2)|\gamma_h(k)\rangle \nn\\
    &+F^{cbe}\langle\jA_\mu^a(q_1) V^e_\nu(p+q_2)|\gamma_h(k)\rangle\,,
\end{align}
where the RHS can be evaluated using the result from the previous section \cref{eq:2-group-vector-result}. As before, from axial current decomposition we can deduce at leading order in soft momenta
\begin{align}
    \langle \pi^c(p) | \jAbar_\mu^a(q_1)\jAbar_\nu^b(q_2)| \gamma_h(k)\rangle = F^{abc}\frac{\kappa}{4\pi^2f_\pi }\epsilon_{\mu\nu\rho\sigma}p^\rho \e^{*\sigma}_h +\cO(q_1,q_2)\,.
    \label{eq:axial-axial_2-group_result}
\end{align}
Therefore we have the 2-group soft axial-axial remnant theorem
\begin{align}
    \lim_{q_1q_2\rightarrow0}\langle \alpha| \jAbar ^a_\mu(q_1)\jAbar^b_\nu(q_2)\rangle =&-i\sum_{i=1}^{n_\pi}(F^{aa_ie}f_X^{ebd}+F^{ba_ie}f_X^{ead}) \frac{\eta_{\mu\nu}}{2p_i\cdot (q_1+q_2)}\cA^{a_1\ldots d\ldots a_n}_{(n_\pi,n_\gamma)} \nn\\
    &-\frac{\kappa}{4\pi^2f_\pi}\sum_{i=1}^{n_\pi}F^{aba_i}\frac{\epsilon_{\mu\nu\rho\sigma}p_i^\rho \e^{*\sigma}_i}{2p_i\cdot (q_1+q_2)} \cA^{a_1\ldots a_{i-1}a_{i+1}\ldots a_n}_{(n_\pi-1,n_\gamma+1)} \nn\\
    &+\frac{\kappa}{4\pi^2f_\pi}\sum_{j=1}^{n_\gamma} F^{abd}\frac{\epsilon_{\mu\nu\rho\sigma}k_j^\rho \e_j^{\sigma}}{2k_j\cdot (q_1+q_2)} \cA^{da_1\ldots a_n}_{(n_\pi+1,n_\gamma-1)}
    \label{eq:soft-axial-axial-theorem-2-group}\,,
\end{align}
where again $\langle \alpha|$ denotes a state with $n_\pi$ pions and $n_\gamma$ photons. The first line of \cref{eq:soft-axial-axial-theorem-2-group} is the contribution from the $G_L\times G_R$ symmetry as derived in Appendix \ref{app:soft-axial-nlsm}. 

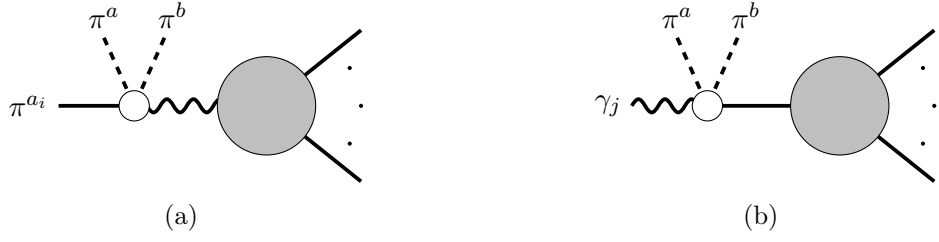
\begin{figure}[t!]
    \centering
    \begin{subfigure}[t]{0.49\textwidth}
        \centering
         \begin{tikzpicture}
            \coordinate (a1) at (-2.4, 0.9);
            \coordinate (a2) at (-1.6, 0.9);
            \node[above]  at (a1) {$\pi^a$};
            \node[above]  at (a2) {$\;\;\pi^b$};
            \coordinate (c) at (-3, 0);	
	    	\coordinate (a) at (-2, 1);
	        \node[left]  at (c) {$\pi^{a_{i}}$};
	    	
	    	\coordinate (f) at (1, 1);
	    	\coordinate (l) at (1, -1);
	    	
	    	\coordinate (m4) at (-2,0);
	    	
	    	\coordinate (mn) at (-0.25, 0);
	    
	    	\coordinate (d1) at (1.00, 0);
	    	\coordinate (d2) at (0.85, 0.5);
	    	\coordinate (d3) at (0.85, -0.5);

                \draw [dashed,hard] (a1) -- (m4);
	    	\draw [dashed,hard] (a2) -- (m4);
            
	    	\draw [hard] (c) -- (m4);
	    	\draw [decorate,decoration={coil,aspect=0},line width=1.4pt] (mn) -- (m4);
	    	
	    	\draw [hard] (f) -- (mn);
	        \draw [hard] (l) -- (mn);
	    	
            \draw[fill=lightgray, opacity=1] (mn) circle (0.65);
            
            \draw[fill=white, opacity=1] (m4) circle (0.2);
	    	
            \draw[fill=black, opacity=1] (d1) circle (0.02);
            \draw[fill=black, opacity=1] (d2) circle (0.02);
            \draw[fill=black, opacity=1] (d3) circle (0.02);
        \end{tikzpicture}
        \caption{}
    \end{subfigure}%
    ~ 
    \begin{subfigure}[t]{0.49\textwidth}
        \centering
        \begin{tikzpicture}
            \coordinate (a1) at (-2.4, 0.9);
            \coordinate (a2) at (-1.6, 0.9);
            \node[above]  at (a1) {$\pi^{a}$};
            \node[above]  at (a2) {$\;\;\pi^{b}$};
            \coordinate (c) at (-3, 0);	
	    	\coordinate (a) at (-2, 1);
	        \node[left]  at (c) {$\gamma_j$};
	    	
	    	\coordinate (f) at (1, 1);
	    	\coordinate (l) at (1, -1);
	    	
	    	\coordinate (m4) at (-2,0);
	    	
	    	\coordinate (mn) at (-0.25, 0);
	    
	    	\coordinate (d1) at (1.00, 0);
	    	\coordinate (d2) at (0.85, 0.5);
	    	\coordinate (d3) at (0.85, -0.5);

                \draw [dashed,hard] (a1) -- (m4);
	    	\draw [dashed,hard] (a2) -- (m4);
            
	    	\draw [decorate,decoration={coil,aspect=0},line width=1.4pt] (c) -- (m4);
	    	\draw [hard] (mn) -- (m4);
	    	
	    	\draw [hard] (f) -- (mn);
	        \draw [hard] (l) -- (mn);
	    	
            \draw[fill=lightgray, opacity=1] (mn) circle (0.65);
            
            \draw[fill=white, opacity=1] (m4) circle (0.2);
	    	
            \draw[fill=black, opacity=1] (d1) circle (0.02);
            \draw[fill=black, opacity=1] (d2) circle (0.02);
            \draw[fill=black, opacity=1] (d3) circle (0.02);
        \end{tikzpicture}
        \caption{}
    \end{subfigure}
    \caption{Graphical representation of the singular contributions comprising the 2-group soft factor $S_\kappa^{(1)}$ in \cref{eq:double-soft-theorem-2-group-schematic} and \cref{eq:double-soft-next-to-leading-order-2-group}.}
    \label{fig:double-soft}
\end{figure}

Finally, plugging in \cref{eq:soft-vector-current-theorem-2-group} and \cref{eq:soft-axial-axial-theorem-2-group} into \cref{eq:double-soft-correlators-in-2-group-section}, we obtain the full double soft theorem \cref{eq:double-soft-theorem-2-group-schematic}. The 2-group contributions in \cref{eq:double-soft-next-to-leading-order-2-group} are represented graphically in \cref{fig:double-soft}.

\subsection{Examples}
Here we provide some examples of tree-level amplitudes in the theory in \Cref{eq:lagrangian-2-group}. We have verified the double soft theorem in all amplitudes with up to six external particles, with any allowed combinations of pions and photons. We only present explicit checks in some simple cases. 

The $\widetilde{C}$ symmetry of \cref{eq: discrete_symmetries} implies a selection rule forbidding all amplitudes with an odd number of external particles. Hence, the first nontrivial amplitudes appear at 4-point. At tree-level we have only $\mathcal{A}_{(4,0)}$ and $\mathcal{A}_{(3,1)}$. The four pion amplitude is the same as in the NLSM \cref{eq:lagrangian-nlsm} for spontaneous 0-form symmetry breaking. The amplitude with three pions (with momenta $p_1, p_2,p_3$) and a photon (momentum $p_4$) is
\begin{align}
\cA^{a_1a_2a_3}_{(3,1)}(p_1,p_2,p_3,p_4)=i\constK f^{a_1a_2a_3}\epsilon(p_1p_2p_4\e_4)\,, \quad \quad \constK=\frac{\kappa}{2\pi^2f_\pi^3}\,,
\label{eq:amp-3-1}
\end{align}
where we used $\constK$ to denote a common combination of constants.

To illustrate the double soft theorem in a simple example, we consider an amplitude with four pions (momenta $p_1,\ldots, p_4$) and two photons (momenta $p_5$ and $p_6$) given by
\begin{align}
\cA^{a_1a_2a_3a_4}_{(4,2)}=&f^{a_1a_2c}f^{ca_3a_4}\constK^2\Bigg[\frac{\epsilon(p_1p_2p_5\e_5)\epsilon(P_{125}p_3p_6\e_6)}{s_{12}+s_{15}+s_{25}}+\frac{\epsilon(p_1p_2p_6\e_6)\epsilon(P_{126}p_3p_5\e_5)}{s_{12}+s_{16}+s_{26}}\Bigg]\nn\\
&+(1\leftrightarrow3)+(2\leftrightarrow3)\,,
\end{align}
where $P^\mu_{ijk}=p^\mu_i+p^\mu_j+p^\mu_k$ and we introduced Mandelstam variables $s_{ij}=2p_i\cdot p_j$.

Taking the double soft limit of two pions, $p_1\to 0$ and $p_2\to0$, we obtain
\begin{align}
\label{eq:amp-4-2-soft}
\lim_{p_1,p_2\to0}\cA^{a_1a_2a_3a_4}_{(4,2)}=&f^{a_1a_2c}f^{ca_3a_4}\constK^2\Bigg[\frac{\epsilon(p_1p_2p_5\e_5)\epsilon(p_5p_3p_6\e_6)}{s_{15}+s_{25}}+\frac{\epsilon(p_1p_2p_6\e_6)\epsilon(p_6p_3p_5\e_5)}{s_{16}+s_{26}}\Bigg]\,,
\end{align}
which is at next-to-leading order $\cO(p_i)$ in soft momenta.

Now we evaluate the double soft limit of $\cA_{(4,2)}$ using the soft theorem in \cref{eq:double-soft-theorem-2-group-schematic}. Note that since $\cA_{(2,2)}$ and $\cA_{(1,3)}$ vanish, we only get non-trivial contributions from the soft operator in \cref{eq:double-soft-next-to-leading-order-2-group} acting on external photons
\begin{align}
   S^{(1)}_{\kappa}\cA_{(2,2)}&=-i\constK f^{a_1a_2 c}\Bigg[\frac{ \epsilon(q_1 q_2 p_5 \e_5)}{s_{15}+s_{25}}\mathcal{A}^{ca_3a_4}_{(3,1)}(p_5,p_3,p_4,p_6)+\frac{ \epsilon(q_1 q_2 p_6 \e_6)}{s_{16}+s_{26}}\mathcal{A}^{ca_3a_4}_{(3,1)}(p_6,p_3,p_4,p_5)\Bigg]\nn\\
   &=  \constK^2f^{a_1a_2 c}f^{ca_3a_4}\Bigg[\frac{ \epsilon(q_1 q_2 p_5 \e_5)}{s_{15}+s_{25}}\epsilon(p_5p_3p_6\e_6)+\frac{ \epsilon(q_1 q_2 p_6 \e_6)}{s_{16}+s_{26}}\epsilon(p_6p_3p_5\e_5)\Bigg]\,,
   \label{eq: soft_factor_on_A_2_2}
\end{align}
where in the second line we plugged in for the amplitudes using \cref{eq:amp-3-1}. This recovers the soft limit obtained in \cref{eq:amp-4-2-soft}, showing that the double soft theorem is satisfied. Note that in the first line of \cref{eq: soft_factor_on_A_2_2} the kinematic pre-factor multiplying each amplitude is odd under parity, $P_0$. However, the sub-leading soft theorem as a whole respects the $P=P_0(-1)^{n_\pi}$ symmetry since $\mathcal{A}_{(3,1)}$ has three external pions.

Next, consider the 6-point pion amplitude, which decomposes into a NLSM part and $\cA_{(6,0)}^{\kappa}$ denoting terms with two separate color-structures
\begin{align}
    \cA_{(6,0)}  =\cA_{(6,0)}^{\textrm{NLSM}}+\cA_{(6,0)}^{\kappa}\,.
    \label{eq:amp-6-0}
\end{align}
The contribution $\cA_{(6,0)}^{\kappa}$ is a sum over ten factorization channels with different color-structures. Writing the $(123)(456)$ channel explicitly, we have
\begin{align}
    \cA_{(6,0)}^\kappa  =\frac{f^{a_1a_2a_3}f^{a_4a_5a_6}\constK^2}{8(s_{12}+s_{13}+s_{23})}\Big[&s_{34}(-s_{15}s_{23}+s_{13}s_{25})+s_{35}(-s_{13}s_{24}+s_{14}s_{23})\nn\\
    &+s_{12}s_{34}(s_{25}-s_{15})+s_{12}s_{35}(s_{14}-s_{24})\nn\\
    &+(s_{13}+s_{23})(s_{15}s_{24}-s_{14}s_{25})\Big]+\cdots\,,
    \label{eq:amp-6-0-k}
\end{align}
where the dots denote the other nine channels, which can be obtained by permutation.

Now we consider the double soft limit of the six pion amplitude in \cref{eq:amp-6-0} and keep the terms to sub-leading order. The $\cA_{(6,0)}^{\textrm{NLSM}}$ satisfies the NLSM double soft theorem; here we focus on $\cA_{(6,0)}^{\kappa}$. Clearly, the double soft limit of \cref{eq:amp-6-0-k} vanishes at leading order in soft momenta $\cO(q^0)$. At next-to-leading order, only the terms with a soft pole survive, and so four channels contribute. Again, focusing on the $(123)(456)$ channel, the double soft limit yields
\begin{align}
\lim_{p_1,p_2 \to 0}\cA_{(6,0)}^{\kappa}  =&
\frac18\frac{f^{a_1a_2a_3}f^{a_4a_5a_6}\constK^2}{s_{13}+s_{23}}\Big[s_{34}(-s_{15}s_{23}+s_{13}s_{25})+s_{35}(-s_{13}s_{24}+s_{14}s_{23})\Big]\nn\\
&+(3\leftrightarrow4)+(3\leftrightarrow5)+(3\leftrightarrow6)\,,
\label{eq:soft-6-0-LHS}
\end{align}
where we used that the second and third lines in \cref{eq:amp-6-0-k} are higher order in soft momenta.

On the other hand, we can evaluate the new contributions to the double soft limit using the soft theorem in \cref{eq:double-soft-next-to-leading-order-2-group}, which is a sum of soft operator acting on all external particles. For instance, when the soft operator acts on the pion with momentum $p_3$, we have
\begin{align}
S^{(1)}_\kappa(p_3)\cA_{(4,0)}  =&
i\constK\sum_{\textrm{spins}}\frac{f^{a_1a_2a_3}\epsilon(p_1p_2p_3\e_3^*)}{s_{13}+s_{23}}\cA^{a_4a_5a_6}_{(3,1)}(p_4,p_5,p_6,p_3)\,,
\end{align}
which we can evaluate using \cref{eq:amp-3-1}. Hence we obtain
\begin{align}
S^{(1)}_\kappa(p_3)\cA_{(4,0)}  
=&-\constK^2\sum_{\textrm{spins}}\frac{f^{a_1a_2a_3}\epsilon(p_1p_2p_3\e_3^*)}{s_{13}+s_{23}}f^{a_4a_5a_6}\epsilon(p_4p_5p_3e_3)\nn\\
=&-\constK^2\frac{f^{a_1a_2a_3}f^{a_4a_5a_6}}{s_{13}+s_{23}}\frac{1}{8}\Big[s_{34}(s_{15}s_{23}-s_{13}s_{25})+s_{35}(s_{13}s_{24}-s_{14}s_{23})\Big]\,,
\end{align}
which agrees with \cref{eq:soft-6-0-LHS}. Repeating the same steps for the other three channels, we see explicitly that the double soft theorem holds for the 6-point pion amplitude.
\section{Conclusions}
In this paper, we explored the implications of spontaneously broken higher symmetries for the soft behavior of scattering amplitudes.  Our principal result was the derivation of a new double soft theorem for NGB in theories possessing a spontaneously broken continuous 2-group global symmetry. This structure is characterized by a current algebra which mixes 0-form and 1-form symmetry currents. We showed that the corresponding Ward identities imply a universal subleading soft factor acting on lower point amplitudes. This soft factor contains terms where particles in the lower point amplitude change species, from the 0-form NGB pion to the 1-form NGB photon, and vice versa. We have also illustrated the new soft theorem with explicit examples of amplitudes in theories with such symmetry.

Along the way, we presented a unified picture for soft theorems from spontaneously broken symmetries. This allowed us to recast well-known results, such as the leading-order soft photon theorem as a consequence of $1$-form symmetry emergent in the soft limit. In this limit, the energy of a soft NGB is much smaller than the energies of massive particles. The leading soft behavior of the amplitude is then captured by an EFT with Wilson lines of charged particles treated as background insertions, similar to those familiar from particle physics such as HQET \cite{Manohar:2000dt}, SCET \cite{Bauer:2000yr}, and their relatives. Our analysis shows that higher-form symmetries can generically emerge as accidental symmetries of such EFTs. It would be interesting explore the connections between this observation to factorization phenomena, and derive the associated soft theorems or selection rules on matrix elements. We expect that such an EFT perspective will also be useful in extending our analysis to sub-leading soft theorems.

Throughout the paper, we carried out the derivations of all soft theorems without making reference to diagrammatics or perturbation theory, and instead using the Ward identities of currents which take the form of ordinary local operators in spacetime. We have not attempted to connect our analysis to the derivation of soft theorems from asymptotic symmetries \cite{Strominger:2017zoo, Campiglia:2015qka, He:2019jjk}. This seems a worthwhile exercise, which is left for the interested reader.

While our analysis of soft theorems focused on soft pions and photons, we expect that it will extend to other interesting cases.
For instance, it is natural to consider the free gluons of non-abelian gauge theories at weak coupling as the (pseudo-)NGB of emergent 1-form symmetries in the zero-coupling limit, around which point their scattering amplitudes are well-defined (see \cite{Cordova:2022rer,Antinucci:2022eat} for related observations). Hence, one can likely derive their soft theorems using analogous symmetry arguments. One might also wonder if a similar picture holds for soft gravitons, perhaps in connection to the symmetries in Refs.~\cite{Benedetti:2021lxj,Hinterbichler:2022agn,Benedetti:2023ipt,Cheung:2024ypq}.

Finally, we believe that further soft theorems might be discovered using higher symmetry as a guiding principle, perhaps in connection to higher-rank or more exotic symmetries in other dimensions, as well as non-invertible symmetries \cite{Chang:2018iay, Cordova:2022fhg,Choi:2022jqy,Damia:2022seq,Bhardwaj:2022yxj,Kaidi:2022uux,Kaidi:2021xfk,Choi:2022zal,Choi:2021kmx,Cordova:2024ypu} which might be related to the existence of massless (or light) particles. We leave such explorations for another time.

\noindent\textbf{Acknowledgements.}
We are grateful to Clifford Cheung, Clay C\'ordova, Thomas Dumitrescu, Ira Rothstein and Tadakatsu Sakai for discussions, and especially  Giovanni Rizi for many useful conversations. We also thank Andreas Helset and Giovanni Rizi for comments on the manuscript. The present research was partially supported by the 2021 Balzan Prize for Gravitation: Physical and Astrophysical Aspects, awarded to T. Damour. J.B.D. is supported by a Four-Year Fellowship (4YF) from the University of British Columbia. M.D. is supported by the DOE under grant no. DE-SC0011632 and by the Walter
Burke Institute for Theoretical Physics.

\appendix 
\section{Select form factor soft limits}
\label{app:select-form-factors-0-form}
In this appendix we provide the technical details for the derivation of the soft behavior of the correlators $\langle\alpha| V^c_\mu(q) \rangle$ and $\langle\alpha|  \jAbar^a_\mu(q_1)\jAbar^b_\nu(q_2)\rangle$ which appeared in \cref{eq:double-soft-correlators} of the main text.
\subsection{Soft vector current}
\label{app:soft-vector-current}
Consider the behavior of the following matrix element where an off-shell vector current is inserted into a scattering amplitude of on-shell NGB,
\begin{align}
    \langle\pi^{a_1}(p_1)\ldots\pi^{a_i}(p_i)\ldots\pi^{a_n}(p_n)| V^a_\mu(q)\rangle. \label{eq:soft-vector-current-matrix-element}
\end{align}
In \cite{Kampf:2013vha} the leading order soft behavior of \cref{eq:soft-vector-current-matrix-element} was proven. Here we will extend that proof to the sub-leading order in the soft momentum $q$. This extension will closely follow the development appearing in \cite{Bern:2014vva}, wherein sub-leading soft theorems are proven for photons, gluons and gravitons.

In the soft limit, the singular behavior of \cref{eq:soft-vector-current-matrix-element} will derive from processes in which a propagating single particle state is created. Such a pole structure will be generated for each of the outgoing NGB states $\langle \pi^{a_i}(p_i)|$. We insert a complete set of states $\mathbbm{1} = \sumint_X |X\rangle \langle X|$ between the soft current operator $V^a_\mu(q)$ and the rest of the process,
\begin{align}
    &\sumint_X\sum^n_{i=1}\langle \pi^{a_i}(p_i)|V^a_\mu(q)|X\rangle\Delta_X\langle X+ \pi^{a_1}(p_1)\ldots\widehat{\pi^{a_i}(p_i)}\ldots\pi^{a_n}(p_n)|0\rangle \\
    &=\sum^n_{i=1}\langle \pi^{a_i}(p_i)|V^a_\mu(q)|\pi^{c}(p_i+q)\rangle\Delta^{cd}(p_i+q)\langle \pi^{a_1}(p_1)\ldots\pi^{d}(p_i+q)\ldots\pi^{a_n}(p_n)|0\rangle\nn
\end{align}
with $\widehat{\pi^{a_i}}$ denoting the omission of the corresponding particle. Where $|X\rangle$ is a multi-particle state, the kinematics in the soft limit will not produce singular behavior. The singular behavior will receive contributions from only those $|X\rangle$ which are single-particle NGB states, where $\Delta^{cd}(p_i+q)$ is the associated propagator.

To obtain the sub-leading contribution we begin from the decomposition,
\begin{align}
    &\lim_{q\to0}  \langle\pi^{a_1}(p_1)\ldots\pi^{a_i}(p_i)\ldots\pi^{a_n}(p_n)| V^a_\mu(q)\rangle\nn\\
    &=-\sum_{i=1}^nf_X^{aa_i d} \frac{(2p_i + q)_\mu}{(p_i + q)^2}\langle\pi^{a_1}(p_1)\ldots\pi^{d}(p_i+q)\ldots\pi^{a_n}(p_n)|0\rangle + R^{aa_1\ldots a_n}_\mu(q;p_1,\ldots,p_n). \label{eq:soft-vector-current-decomposition}
\end{align}
The leading order behavior follows from a derivation analogous to that in Appendix E of \cite{Kampf:2013vha}. The remnant term, $R^{aa_1\ldots a_n}_\mu$, parameterizes insertions of the current which do not contain the $(p_i+q)$ pole. The entire matrix element satisfies the Ward identity,
\begin{align}
    0&=q^\mu \langle\pi^{a_1}\cdots\pi^{a_n}| V^a_\mu(q)\rangle\nn\\
    &= \sum_{i=1}^nf_X^{aa_i d} \langle\pi^{a_1}(p_1)\ldots\pi^{d}(p_i+q)\ldots\pi^{a_n}(p_n)|0\rangle+q^\mu R^{aa_1\ldots a_n}_\mu(q;p_1,\ldots,p_n).
\end{align}
Expanding around $q=0$, we have
\begin{align}
    0&= \sum_{i=1}^nf_X^{aa_i d}\left(1 + q^\mu\frac{\partial}{\partial p_i^\mu}\right)\langle\pi^{a_1}(p_1)\ldots\pi^{d}(p_i)\ldots\pi^{a_n}(p_n)|0\rangle\nn\\
    &+ q^\mu R^{aa_1\ldots a_n}_\mu(0;p_1,\ldots,p_n) + \mathcal{O}(q^2).
\end{align}
This produces a set of relations for each order in $q$. At leading order we have,
\begin{align}
    0&= \sum_{i=1}^nf_X^{aa_i d} \langle\pi^{a_1}(p_1)\ldots\pi^{d}(p_i)\ldots\pi^{a_n}(p_n)|0\rangle,
\end{align}
which is a consequence of invariance under the unbroken subgroup $H$, and was demonstrated in \cite{Kampf:2013vha}. At the next order we have
\begin{align}
    \sum_{i=1}^nf_X^{aa_i d}q^\mu\frac{\partial}{\partial p_i^\mu}\langle\pi^{a_1}(p_1)\ldots\pi^{d}(p_i)\ldots\pi^{a_n}(p_n)|0\rangle = -q^\mu R^{aa_1\ldots a_n}_\mu(0;p_1,\ldots,p_n). \label{eq:soft-vector-current-NLO-relation}
\end{align}
This relation determines the remnant $R^{aa_1\ldots a_n}_\mu(0;p_1,\ldots,p_n)$ up to potential terms that are separately vanishing under the Ward identity. Furthermore, the requirement that such terms be local in $q$ implies that they must also be at least linear in $q$. Therefore these terms would be suppressed in the soft expansion. We can remove the $q^\mu$ contracted with both sides of \cref{eq:soft-vector-current-NLO-relation}, leaving
\begin{align}
    \sum_{i=1}^nf_X^{aa_i d} \frac{\partial}{\partial p_i^\mu}\langle\pi^{a_1}(p_1)\ldots\pi^{d}(p_i)\ldots\pi^{a_n}(p_n)|0\rangle = - R^{aa_1\ldots a_n}_\mu(0;p_1,\ldots,p_n). 
\end{align}
Inserting this into the decomposition \cref{eq:soft-vector-current-decomposition} completes the sub-leading theorem,
\begin{align}
&\lim_{q\rightarrow 0}\langle\pi^{a_1}\cdots\pi^{a_n}| V^a_\mu(q)\rangle\nn \nonumber\\
&=-i\sum_{i=1}^n f^{a a_i d}\left(\frac{(2p_{i}+q)_\mu}{(p_i+ q)^2} -\frac{iq^\nu \angMom_{i\mu\nu}}{(p_i\cdot q)} \right)\cA_{n}^{a_1\ldots d\ldots a_n}(p_1,\ldots,p_n)
+ \mathcal{O}(q) \label{eq:soft-vector-current-theorem}
\end{align}
where as before $\angMom_{i\mu\nu}$ is the angular momentum operator defined as
\begin{align}
     \angMom_i^{\mu\nu} = i \left(p_i^\mu\frac{\partial}{\partial p_{i\nu}}-p_i^\nu\frac{\partial}{\partial p_{i\mu}}\right).
\end{align}

\subsection{Soft axial-axial remnant}
\label{app:soft-axial-nlsm}
We will proceed with a similar analysis for the matrix element:
\begin{align}
    \langle\pi^{a_1}(p_1)\ldots\pi^{a_i}(p_i)\ldots\pi^{a_n}(p_n)| \jAbar ^a_\mu(q_1)\jAbar ^b_\nu(q_2)\rangle. \label{eq:soft-axial-axial-current-matrix-element}
\end{align}
We will follow the same line of reasoning as for the soft vector current $V^a_\mu(q)$ in the matrix element \cref{eq:soft-axial-axial-current-matrix-element}. The singular behavior of \cref{eq:soft-axial-axial-current-matrix-element} will derive from contributions of the form,
\begin{align}
    \sum^n_{i=1}\langle \pi^{a_i}(p_i)|&\jAbar ^a_\mu(q_1)\jAbar ^b_\nu(q_2)|\pi^{c}(p_i+q_1+q_2)\rangle\nn\\
    &\times\Delta^{cd}(p_i+q_1+q_2)\langle \pi^{a_1}(p_1)\ldots\pi^{d}(p_i+q_1+q_2)\ldots\pi^{a_n}(p_n)|0\rangle,
\end{align}
which are generated by the propagation of single particle NGB states. 

The isolated form factor $\langle \pi^{a_i}(p_i)|\jAbar ^a_\mu(q_1)\jAbar ^b_\nu(q_2)|\pi^{c}(p_i+q_1+q_2)\rangle $ can be determined to leading order. 
The effective theory is invariant under parity inversions, thus we consider a parity-even ansatz,
\begin{align}
    &\langle \pi^{a_i}(p_i) | \jAbar_\mu^a(q_1)\jAbar_\nu^b(q_2)|\pi^c(p_i+q_1+q_2)\rangle\nn\\ &= A_1(p_i,q_1,q_2)B_1^{a_iabc}\eta_{\mu\nu} +  A_2(p_i,q_1,q_2)B_2^{a_iabc}p_{i\mu}p_{i\nu}+\mathcal{O}(q_1,q_2), \label{eq:pi-A-A-pi-ansatz}
\end{align}
where $A_i(p_i,q_1,q_2)$ are Lorentz-invariant structure functions and $B_i^{a_iabc}$ are arbitrary flavor-structures. The $\jAbar^a_\mu$ operators appearing in this form factor do not satisfy any Ward identities, but we can instead consider the Ward identity,
\begin{align}
    &p_i^\alpha\langle  \jA^{a_i}_\alpha(p_i)\jA_\mu^a(q_1)\jA_\nu^b(q_2)|\pi^c(p_i+q_1+q_2)\rangle\label{eq:ward-hard-three}\\
    &=F^{a_iad}\langle V^d_\mu(p_i+q_1)\jA_\nu^b(q_2)|\pi^c(p_i+q_1+q_2)\rangle +F^{a_ibd}\langle \jA_\mu^a(q_1)V^d_\nu(p_i+q_2)|\pi^c(p_i+q_1+q_2)\rangle\,.\nn 
\end{align}
Through a procedure of writing analogous ans\"atze for the correlators on the RHS of the Ward identity \cref{eq:ward-hard-three}, imposing the subsequent Ward identities these correlators in turn satisfy, and finally matching the contained pole structures with our initial ansatz \cref{eq:pi-A-A-pi-ansatz}, we arrive at
\begin{align}
    &\langle \pi^{a_i}(p_i) | \jAbar_\mu^a(q_1)\jAbar_\nu^b(q_2)|\pi^c(p_i+q_1+q_2)\rangle\nn \\
    &= -i(F^{a_iae}f_X^{ebc}+ F^{a_ibe}f_X^{eac})\eta_{\mu\nu} +  A_2(p_i,q_1,q_2)B_2^{a_iabc}p_{i\mu}p_{i\nu}+\mathcal{O}(q_1,q_2)\,.
    \label{eq:axial_axial_nlsm_result}
\end{align}
The undetermined product $A_2(p_i,q_1,q_2)B_2^{a_iabc}$ is sensitive to higher-derivative operators in the effective field theory, which was studied in \cite{Rodina:2021isd}, and also one loop corrections. For simplicity we have suppressed these corrections in the sub-leading NLSM soft factor \cref{eq:NLSM-double-soft-subleading-order}. These same corrections are also implicitly suppressed in the sub-leading 2-group soft theorem \cref{eq:double-soft-theorem-2-group-schematic}, where they would be contained in the NLSM contribution $S^{(1)}$. 

At tree-level, we can conclude the soft behavior,
\begin{align}
    &\lim_{q_1q_2\rightarrow0}\langle \alpha| \jAbar^a_\mu(q_1)\jAbar^b_\nu(q_2)\rangle\nn\\
    &= -i\sum_{i=1}^n(F^{aa_ie}f_X^{ebd}+F^{ba_ie}f_X^{ead}) \frac{\eta_{\mu\nu}}{2p_i\cdot (q_1+q_2)}\cA^{a_1\cdots d\cdots a_n}_n
    +\mathcal{O}(q^0)\, .\label{eq:soft-remnant-NLSM}
\end{align}
Combining the results \cref{eq:soft-remnant-NLSM} and \cref{eq:soft-vector-current-theorem} in \cref{eq:double-soft-correlators}, we arrive at the NLSM double soft theorem \cref{eq:double_soft_nlsm} in the main text.

\bibliographystyle{JHEP}
\bibliography{references}

\end{document}